\documentclass[11pt]{article}
\pdfoutput=1

\usepackage{jheppub}

\usepackage{verbatim} 
\usepackage{float}
\restylefloat{table}
\usepackage[all]{xy}
\usepackage{pdflscape}

\usepackage{amsmath,amssymb,epsfig,amsfonts,csquotes,multirow}
\usepackage{tikz,subcaption,booktabs}
\usetikzlibrary{intersections,calc,fit,shapes}
\usetikzlibrary[decorations.pathmorphing]
\newcommand{\cd}{\mathrm{CxDef}}
\newcommand{\kd}{\mathrm{KaDef}}
\newcommand{\ib}[1]{\{ #1 \}}

\usepackage{array}
\usepackage{youngtab}
\usepackage{multirow}
\usepackage{color}
\usepackage{ulem}

\makeatletter

\newcommand{\tw}[1]{{\color{red} #1}}

\newcommand{\be}{\begin{equation}}
\newcommand{\ee}{\end{equation}}
\newcommand{\ba}{\begin{aligned}}
\newcommand{\ea}{\end{aligned}}

\newcommand{\bea}{\begin{eqnarray}}
\newcommand{\eea}{\end{eqnarray}}

\newcommand{\ki}{\chi_{\text{top}}}
\newcommand{\p}{\mathbb{P}^1}

\def\unit{{1\kern-.65ex {\rm l}}}
\def\1{{1\kern-.65ex {\rm l}}}

\def\bbP{\mathbb{P}}

\def\tw{{\rm w}}

\def\bbN{{\mathbb{N}}}

\def\bbP{{\mathbb{P}}}
\def\bbQ{{\mathbb{Q}}}

\newcount\hour \newcount\minute
\hour=\time \divide \hour by 60
\minute=\time
\def\now{%
\ifnum \hour<13
  \ifnum \hour=0 \advance \hour by 12 \number\hour:\else \number\hour:\fi%
     \ifnum \minute<10 0\fi%
     \number\minute%
\ A.M.%
\else \advance \hour by -12 \number\hour:%
  \ifnum \minute<10 0\fi%
  \number\minute%
  \ P.M.%
\fi%
}

\makeatother

\begin{document}

\numberwithin{equation}{section}  



\vspace*{-2cm} 
\begin{flushright}
\end{flushright}

\vspace*{0.8cm} 
\begin{center}
 {\LARGE  Terminal Singularities, Milnor Numbers, \vspace{3mm}\\and Matter in F-theory 
}
 
 \vspace*{1.8cm}
{Philipp Arras$\, ^1$, Antonella Grassi$\,^2$, and Timo Weigand$\,^1$}\\

 \vspace*{1.0cm} 
\smallskip
{\it $^1$ Institut f\"ur Theoretische Physik, Ruprecht-Karls-Universit\"at,\\
 Philosophenweg 19, 69120 Heidelberg, Germany }\\
  {\tt {email:$\,$ p. arras, t.weigand\phantom{@}thphys.uni-heidelberg.de}}

{\it $^2$ Department for Mathematics, University of Pennsylvania,\\
209 S 33rd Street, Philadelphia, PA 19104, USA }\\
  {\tt {email:$\,$ grassi \phantom{@}upenn.edu}}

\vspace*{0.8cm}
\end{center}
\vspace*{.5cm}

\noindent

We initiate a systematic investigation of F-theory on elliptic fibrations with singularities which  cannot be resolved without breaking the Calabi-Yau condition, corresponding to $\bbQ$-factorial terminal singularities.
It is the purpose of this paper to elucidate the physical origin of such  non-crepant  singularities in codimension two and to systematically analyse F-theory compactifications containing such singularities. 
The singularities reflect the presence of localised matter states from wrapped M2-branes which are not charged under any massless gauge potential.
We identify a class of $\mathbb Q$-factorial terminal singularities on elliptically fibered Calabi-Yau threefolds for which we can compute the number of uncharged  localised hypermultiplets in terms of their associated Milnor numbers. These count the local complex deformations of the singularities. The resulting six-dimensional spectra are shown to be anomaly-free.  
We exemplify this in a variety of cases, including models with non-perturbative gauge groups with both charged and uncharged localised matter. 
The underlying mathematics will be discussed further in a forthcoming publication.

\newpage
\tableofcontents

\section{Introduction}

This article investigates a certain class of singularities in elliptically fibered Calabi-Yau threefolds which cannot be resolved without breaking the Calabi-Yau condition.
Singularities of this type appear frequently in compactifications of F-theory and require new techniques for the computation of the massless spectrum of the associated effective field theory. Our systematic investigation  of such singularities reveals new contributions to the localised matter spectrum, and we provide the mathematical methods
to compute them.

Indeed, geometric singularities play a distinguished role in compactifications of string and M-theory.
String theory is the ideal framework to study compactification on singular spaces: It contains just the right type of extended BPS objects to oftentimes render the lower-dimensional physics completely well-defined despite the appearance of a singularity from the perspective of classical geometry.
Indeed,
when a singularity arises as  a cycle shrinks to zero volume, wrapped BPS branes become massless and their inclusion spectacularly resolves the seeming singularity in the Wilsonian effective action of the string compactification \cite{Strominger:1995cz}.
Relatedly, geometric singularities typically signal an enhancement of the symmetries governing the effective physics, as they usually sit at the intersection of a Coulomb and Higgs branch of the effective field theory.

This general lore is at the heart of F-theory \cite{Vafa:1996xn,Morrison:1996na,Morrison:1996pp} and its dual formulation via M-theory compactified on an elliptically  fibered Calabi-Yau space.
The latter 
provides a beautiful dictionary between the geometric structure of singularities in the elliptic fiber and the effective physics governing the dynamics of compactifications with 7-branes. 
The traditional way to deal with such singularities is to perform a resolution and to infer the physics associated with the original, singular model by taking a suitable limit.
This procedure works particularly well if the singularities allow for a {\it crepant resolution} of the singular Calabi-Yau, which by definition does not change the canonical bundle of the space. In particular, since the crepant resolution of a singular Calabi-Yau space is still Calabi-Yau, supersymmetry is preserved along the way. 
A crepant resolution of the fibral singularities corresponds to moving along a flat direction in the K\"ahler moduli space by giving a non-zero volume to the vanishing curves in the fiber whose shrinking has created the singularity. In this way zero modes from M2-branes wrapped around the vanishing cycles  become massive. 
In the singular limit, the wrapped M2-branes at singularities in codimension one give rise to non-abelian gauge bosons \cite{Witten:1996qb}, and the resolution hence corresponds to moving away from the origin of the Coulomb branch. Here and in the sequel the Coulomb branch we are referring to is the one of the M-theory compactification, e.g. to $\mathbb R^{1,4}$ for a Calabi-Yau threefold, not of the dual F-theory vacuum in one dimension higher (see also Section \ref{sec:Crepant} for a review). As a result,  one can make an association between the resolved fibers in codimension one and  the affine Dynkin diagrams of A-D-E type.\footnote{In fact in codimension one the Calabi-Yau and the associated Jacobian  are isomorphic and  one considers the fiber components which do not intersect the section of the Jacobian.}

As two such codimension-one strata intersect, or one self-intersects, in codimension two on the base, new vanishing cycles might arise in the fiber. These are in 1-1 correspondence with the weight system of representations of the Lie algebra of the model. This is because M2-branes wrapping the codimension-two fiber curves form charged matter states.
The possible crepant resolutions are in a beautiful match with the different phases of the Coulomb branch \cite{Intriligator:1997pq,Aharony:1997bx,Grimm:2011fx,Hayashi:2013lra,Hayashi:2014kca, Esole:2014bka, Esole:2014hya,Braun:2014kla,Lawrie:2015hia,Braun:2015hkv}. We will briefly review this connection further in Section \ref{sec:Crepant} as it plays a key role for our analysis.\footnote{Singularities in codimension three and codimension four encode cubic \cite{Donagi:2008ca,Beasley:2008dc,Beasley:2008kw} and, respectively, quartic \cite{Schafer-Nameki:2016cfr,Apruzzi:2016iac} Yukawa interactions in F-theory compactifications to four and two spacetime dimensions.} 
There is, however, an important difference between the structure of singularities in codimension one and higher of a singular  elliptic Calabi-Yau as above: The first type always admits a crepant resolution; in the second case, by contrast, the singularity type might defy a resolution which does not break the Calabi-Yau condition.

 \vskip 0.1in
 
In this article we elucidate the physical origin of {\it non-crepant resolvable singularities in codimension two} and find a way to systematically work with F-theory compactifications containing such singularities\footnote{In related work to appear, the second author, with Halverson and Shaneson, will consider a local version of a similar situation using a different method.
}.  We focus on six-dimensional F-theory models on elliptically fibered Calabi-Yau threefolds and their dual M-theory compactifications, whose most important properties relevant to our analysis are summarized in Section \ref{sec_Ftheorynotation}. 
As stressed already, from a physics perspective, a crepant resolution represents a flat direction in the classical Coulomb branch of the dual M-theory along which localised charged matter acquires a mass. Turning tables around, we will argue in Section \ref{subsec_non-crepantb} that a singularity without a crepant resolution must host massless matter from wrapped M2-branes which cannot be rendered massive along any flat direction in Coulomb branch. This means that the localised matter must be uncharged, at least under any massless gauge potential of the compactification. 
Indeed, examples of such behaviour have already appeared in \cite{Braun:2014nva}, in \cite{Braun:2014oya}, which studies the singular Jacobian of a genus one-fibration, and more recently in \cite{Morrison:2016lix}. The relation between singularities and the presence of matter charged only under discrete gauge groups has been stressed further in \cite{Morrison:2014era,Mayrhofer:2014haa,Mayrhofer:2014laa}.  
As we will see in this work, the phenomenon of uncharged localised matter is, however, much more general. 

\begin{figure}
\centering
\begin{tikzpicture}
\node [align=left,draw,ellipse] (A) at (3,2) {{\bf Uncharged hyper multiplets:}\\{ $  \frac{1}{2} ( b_3(X) + \sum m_P)$}};
\node [align=left,draw] (B) at (-1,0) {{\bf Non-localised hypers:} \\{  $\frac{1}{2} ( b_3(X) - \sum_P m_P)$}};
\node [align=left,draw] (C) at (6,0) {{\bf Localised hypers:} \\ { $\sum m_P=\sum \tau_P$}};
\node [draw,circle] (+) at (3,0) {$+$};
\draw [->,very thick] (B) -- (+);
\draw [->,very thick] (C) -- (+);
\draw [->,ultra thick] (+) -- (A);
\end{tikzpicture}
\caption{Origin of uncharged hyper multiplets in six-dimensional F-theory compactifications.}
 \label{fig:unchargedHypers}
\end{figure}
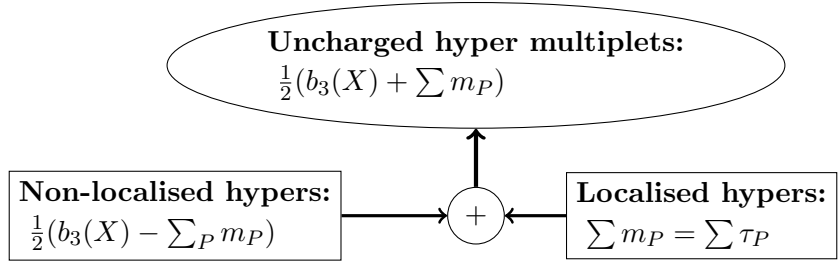

In Section \ref{sec:Fnoncrepant} we  identify a class of ${\mathbb Q}$-factorial terminal singularities, which arise naturally from  Weierstrass models.
Their mathematical properties \cite{ArrasGrassiWeigandM} allow us to deduce the precise number of localised matter states despite the absence of a Calabi-Yau resolution.
The idea is to interpret the localised uncharged states as part of the space of complex structure deformations of the singular variety. 
The latter can be computed by employing its definition via the complex structure moduli space of a nearby deformation.
In this way, we find a formula for the number of localised hypermultiplets in terms of the dimension of the space of  the local  (versal) deformations of each codimension-two singularity. This dimension is the Tyurina-Milnor number of the terminal codimension-two singularity (Section \ref{mathem_background3}). 
   The complex deformations have a natural splitting (summarised in Figure \ref{fig:unchargedHypers}) in terms of the third Betti number $b_3(X)$ of the singular threefold $X$ and the Tyurina numbers of the singularities (Section \ref{UnchLocHyper}). The K\"ahler deformations are computed by $b_2(X)$  as in the smooth case (Section  \ref{sek_defs}, and more generally in \cite{ArrasGrassiWeigandM}), which gives the tensor multiplets.

Another aspect of our analysis is to establish a relation between the topological Euler characteristic and the number of hypermultiplets in Section \ref{sec_defs}.
Note that on singular spaces Poincar\'e  duality and the Hodge decomposition might not hold and half the topological Euler characteristic is not the difference between the K\"ahler and complex structure deformations.  We stress that the modified  relation which we find is the same for all the types of singularities we consider; however, different  arguments are needed in different cases, as we discuss in Section \ref{mathem_background} and further in \cite{ArrasGrassiWeigandM}.
To apply these results concretely,  we review and extend the systematic computation of \cite{Grassi:2000we} of the topological Euler characteristic on Calabi-Yau threefolds in appendix \ref{app-chi}.

We illustrate our general findings in a number of examples of singular elliptic fibrations over a two-dimensional base $B_2$. 
In Section \ref{sec:modelswithoutgauge} we analyze two models with non-trivial gauge group, but with $\mathbb Q$-factorial terminal singularities in codimension two, given by the ${\rm I}_1$ model which has already been studied in \cite{Braun:2014nva,Martucci:2015dxa} as well as a non-perturbative model with enhancements of the form type II $\rightarrow$ III over isolated points. The validity of the spectra we find is checked to satisfy the stringent six-dimensional anomaly cancellation conditions.

While, given the physical explanation sketched above, the lack of a crepant resolution for models without a gauge group might not come as a surprise, we 
exemplify a similar phenomenon for a family of models with non-trivial gauge algebra. Specifically, Section \ref{sec_nontrivialgaugegroup} is dedicated to a family of Weierstrass fibrations with codimension-one fibers of type {III}, corresponding to gauge group $SU(2)$. As we will see, the singularities in codimension-two can become precisely of $\mathbb Q$-factorial terminal Kleinian type $A_n$ and hence defy a crepant resolution. This implies that at such loci, both charged and uncharged hypermultplets localize. We provide a partial resolution of these geometries and verify consistency of our claims by establishing an anomaly-free spectrum in the six-dimensional F-theory compactification. 
These two classes of models are generalized further in appendix \ref{app:resolutionTwoBranes}. In appendix \ref{app:toricResolutionsOneBrane} we also provide a detailed resolution of a Weierstrass model with singularities of type IV  in codimension-one which, although crepant resolvable, we find interesting in itself. We conclude in Section \ref{sec:Conclusions} with directions for future research.

\section{F-theory on Resolvable Elliptic Threefolds} \label{sec_Ftheorynotation}

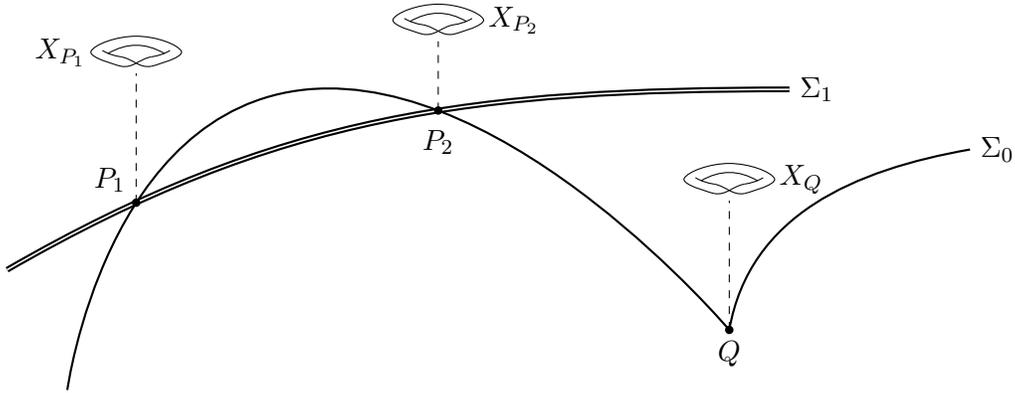
\begin{figure}
\centering

\begin{tikzpicture}[scale=.8,thick]
\def\torus{
	\begin{tikzpicture}[remember picture,scale=.05]
	\coordinate (blub) at (4,0);
	\coordinate (A) at (24,0);
	\coordinate (B) at (12,4);
	\coordinate (C) at (0,0);
	\coordinate (D) at ($(12,-4)-(blub)$);
	\coordinate (E) at ($(12,-4)+(blub)$);
	\coordinate (F) at (12,-3);
	\coordinate (G) at ($(A)+(-3,0.2)$);
	\coordinate (H) at ($(C)+(3,0.2)$);
	\coordinate (X) at (4.5,-.6);
	\coordinate (Y) at (12,1.5);
	\coordinate (Z) at (19.5,-.6);

	\draw plot [smooth,tension=.8] coordinates {(E) (A) (B) (C) (D) };

	\draw (X) to [bend left] (Z);

	\coordinate (G1) at ($(G) + (-5,-1.7)$);
	\coordinate (H1) at ($(H) + (5,-1.7)$);

	\draw (G) to[in=10,out=210] (G1);
	\draw (H) to[in=170,out=-30] (H1);

	\draw (D) to[in=190,out=-10] (G1);
	\draw (E) to[in=-10,out=190] (H1);
	\end{tikzpicture}}

\coordinate (S10) at (2,10);    
\coordinate (S11) at (15,13);  
\coordinate (S00) at (3,8);    
\coordinate (Q) at (14,9);     %
\coordinate (S02) at (18,12);  
\coordinate (V1) at (7,13);    

\draw [name path=Sigma1, double] (S10) to[out=30,in=180] (S11);
\draw [name path=Sigma0] plot [smooth,tension=1.1] coordinates { (S00) (V1) (Q) } ;
\draw (Q) to[out=80,in=190] (S02);
\node [right] at (S11) {$\Sigma_1$};
\node [right] at (S02) {$\Sigma_0$};

\fill [name intersections={of=Sigma0 and Sigma1}]
    (intersection-1) coordinate (P1)
    (intersection-2) coordinate (P2);

\fill[black] (P1) circle (2pt) node[above left] {$P_1$};
\fill[black] (P2) circle (2pt) node[below,yshift=-.3em] {$P_2$};
\fill[black] (Q) circle (2pt) node[below] {$Q$};

\coordinate (XQ) at ($(Q) + (0,2.5)$);
\draw [dashed, thin] (Q) -- ($(XQ) + (0,-.35) $);
\node at (XQ) {\torus};
\node [right=15] at (XQ) {$X_Q$};

\coordinate (XP1) at ($(P1) + (0,2.5)$);
\draw [dashed, thin] (P1) -- ($(XP1) + (0,-.35) $);
\node at (XP1) {\torus};
\node [left=15] at (XP1) {$X_{P_1}$};

\coordinate (XP2) at ($(P2) + (0,1.5)$);
\draw [dashed, thin] (P2) -- ($(XP2) + (0,-.35) $);
\node at (XP2) {\torus};
\node [right=15] at (XP2) {$X_{P_2}$};
\end{tikzpicture}

\caption{Our notation for an elliptically fibered Calabi-Yau manifold with $\Delta = \Sigma_1 \cup \Sigma_0$.}
\label{fig:chi}
\end{figure}

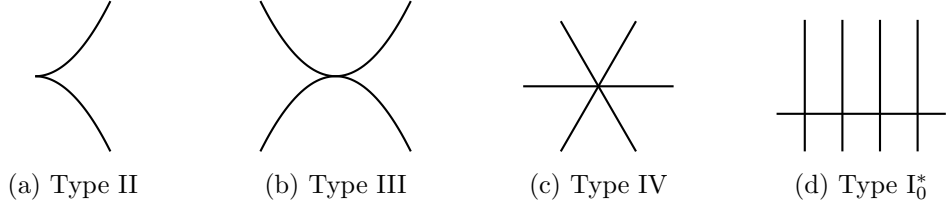
\begin{figure}
\centering
\begin{subfigure}[b]{0.2\textwidth}
	\centering

    \begin{tikzpicture}[thick]
\draw[domain=0:1,smooth,variable=\x] plot ({\x},{\x*\x});
\draw[domain=0:1,smooth,variable=\x] plot ({\x},{-\x*\x});
\end{tikzpicture}
    
    \caption{Type II}
\end{subfigure}
~
\begin{subfigure}[b]{0.2\textwidth}
	\centering

   \begin{tikzpicture}[thick]
\draw[domain=-1:1,smooth,variable=\x] plot ({\x},{\x*\x});
\draw[domain=-1:1,smooth,variable=\x] plot ({\x},{-\x*\x});
\end{tikzpicture}
   
    \caption{Type III}
\end{subfigure}
~
\begin{subfigure}[b]{0.2\textwidth}
	\centering

    \begin{tikzpicture}[thick]
\draw (-1,0) -- (1,0);
\draw[cm={cos(60) ,-sin(60) ,sin(60) ,cos(60), (0,0)}] (-1,0) -- (1,0);
\draw[cm={cos(60) ,sin(60) ,-sin(60) ,cos(60), (0,0)}] (-1,0) -- (1,0);
\end{tikzpicture}

    \caption{Type IV}
\end{subfigure}
~
\begin{subfigure}[b]{0.2\textwidth}
	\centering

    \begin{tikzpicture}[scale=.5,thick]
\coordinate (a) at (1,0);
\coordinate (b) at (2,0);
\coordinate (c) at (3,0);
\coordinate (d) at (4,0);
\coordinate (len) at (0,3.5);
\draw (a) -- ($(a) + (len)$);
\draw (b) -- ($(b) + (len)$);
\draw (c) -- ($(c) + (len)$);
\draw (d) -- ($(d) + (len)$);
\draw (0.25,1) -- (4.75,1);
\end{tikzpicture}
    
    \caption{Type $\mathrm I_0^*$}
\end{subfigure}
\caption{Some Kodaira fiber types appearing in this paper.}
\label{fig:fiberTypes}
\end{figure}

F-theory compactifications to six dimensions are defined in terms of an elliptically fibered\footnote{More generally, one can consider F-theory on torus-fibrations which lack a zero-section \cite{Braun:2014oya}.} Calabi-Yau threefold $Y_3$ with base $B_2$, 
\be
\ba
\pi :\quad \mathbb{E}_\tau \ \rightarrow & \  \ Y_3 \cr 
& \ \ \downarrow \cr 
& \ \  B_2 
\ea
\ee
given in Weierstrass form as a hypersurface $P_W = 0$ with
\begin{align*}
P_W := - y^2 +  x^3+f \,  x^2\, z^4+g \, z^6.
\end{align*}
Here $f,g$ are sections of ${\mathcal O} ( -4 K_B)$ and ${\mathcal O}( -6 K_B)$, respectively, with $K_B$ denoting the canonical bundle of $B_2$. 
The fiber over points on $B_2$ where the discriminant
\bea
\Delta  = 4 \, f^3 + 27 \, g^2
\eea
vanishes is singular, indicating the presence of a 7-brane.
The perhaps simplest non-trivial class of models contains a  specific 7-brane stack along a divisor $\Sigma_1$ parametrised locally by the vanishing of the coordinate  $z_1=0$ on $B_2$, together with a single 7-brane over a divisor $\Sigma_0$, required in order for the 7-brane tadpole to be cancelled.\footnote{For simplicity, throughout this paper we assume that the Mordell-Weil rank of the elliptic fibration is trivial. This assumption does not affect our results and may easily be dropped.} Such a situation, depicted schematically in figure \ref{fig:chi}, is modeled by choosing
\bea \label{eq:fandg}
f = z^{\mu_f}_1  f_0, \quad g= z_1^{\mu_g} g_0 \quad {\rm such\, \, that} \quad  \Delta = z_1^m \, \sigma_0,
\eea 
where $\sigma_0 = 0$ defines the residual discriminant $\Sigma_0$ and $f_0$ and $g_0$ are generic. The Kodaira type of the fiber over generic points on $\Sigma_1$ is determined by the vanishing order $(\mu_f, \mu_g,m)$ of $f$, $g$, and $\Delta$, with two special cases: If $(\mu_f, \mu_g,m) = (0,0,1)$, the fiber is of type ${\rm I}_1$, corresponding to a singular nodal $\mathbb P^1$, and if $(\mu_f, \mu_g,m) = (1,1,2)$, the fiber is a cuspidal $\mathbb P^1$, denoted as fiber type II. In both cases $Y_3$ is smooth as a fibration. In all other cases, the singularity of the fiber is also a singularity of $Y_3$, indicating the presence of  a non-trivial gauge algebra with associated gauge group $G$ on the 7-brane along $\Sigma_1$.\footnote{We will not be very careful distinguishing between the gauge algebra and the gauge group \cite{Mayrhofer:2014opa}. For simplicity we furthermore assume that there are no extra abelian gauge group factors, an assumption which can easily be relaxed.}  In this case, a resolution of $Y_3$ exhibits multi-component fibers classified by their Kodaira type. Some of the fiber types encountered in this article are shown in  figure \ref{fig:fiberTypes}. For the reader's convenience we also include the vanishing orders appearing in Kodaira's classification and the associated gauge algebra in codimension one in table \ref{tab:KodairaTate}.

The singularity type of the fiber enhances further in codimension two, i.e. over points on $B_2$. In the above setup, there are two types of such points: The first corresponds to the intersection  $\Sigma_1 \cap \Sigma_0$ consisting of $B_i$ points of type $P_i$, where the index $i$ labels the fiber type over $P_i$. 
The second type of points are given by the points $Q$ 
where the residual discriminant acquires a cuspidal singularity (as a divisor in $B_2$). Incidentally, over the latter points the fiber type enhances to type II, while for $P_i$ the specific fiber type depends on the vanishing orders of $f$, $g$, and $\Delta$.
Being the intersection between two divisors, points of type $P_i$ carry localised matter in form of massless hypermultiplets.
Our interest in this paper is in particular in this localised matter.

With a few exceptions, discussed in more detail in Section \ref{subsec_non-crepantb}, the literature has focused on singular $Y_3$ which allow for a crepant resolution $\hat Y_3$, i.e. a resolution such that $\hat Y_3$ is itself Calabi-Yau. 
As will be explained in more detail in Section \ref{sec:Crepant}, the existence of a crepant resolution implies that the total gauge group $G$ must be non-trivial, and furthermore that the matter hypermultiplets localised at the intersection points $P_i$ are charged, transforming in some representation $R$ of $G$.

The 6d effective action of F-theory on $Y_3$ is given by an ${\cal N}=(1,0)$ supergravity theory, which is well-known to be subject to a number of non-trivial constraints from the cancellation of gauge and gravitational anomalies. 
Of particular interest for us is the famous condition
\bea \label{6dgravan}
n_H - n_V + 29 \, n_T = 273
\eea
for the cancellation of gravitational anomalies. Here $n_V = {\rm dim}(G)$ is the number of vector multiplets, $n_T = h^{1,1}(B_2) -1$ counts the number of tensor multiplets and $n_H$ counts the total number of hypermultiplets. 
In models with crepant resolution, the origin of the hypermultiplets is one of the following four:
Each point $P_i$ gives rise to a localised, charged hypermultiplet in some representation $R$ of $G$, and from the bulk of $\Sigma_1$ one finds in addition 
$g$ hypermultiplets in some representation of $G$, with $g$ being the genus of $\Sigma_1$. Apart from this charged matter, there is one universal 
hypermultiplet containing the overall volume modulus as well as $h^{2,1}(\hat Y_3)$ uncharged hypermultiplets associated with the complex structure moduli, i.e.
\bea
n_H = n_H^0 + n_H^c, \qquad \quad n_H^0 = 1 + h^{2,1}(\hat Y_3),
\eea
where $n_H^0$ and $n_H^c$ denote the number of neutral and charged hypermultiplets, respectively.

The computation of $h^{2,1}(\hat Y_3)$, and thus of $n_H^0$, is facilitated by the fact that it appears in the topological Euler characteristic $\chi_{\rm top}(\hat Y_3)$ of the smooth resolution $\hat Y_3$, 
\bea
\chi_{\rm top}(\hat Y_3) := \sum_{i=1}^6 (-1)^i b_i(\hat Y_3) = 2 \left(h^{1,1}(\hat Y_3) - h^{2,1}(\hat Y_3) \right) \,.
\eea 
Taking furthermore into account that by the Shioda-Tate-Wazir theorem
\bea
h^{1,1}(\hat Y_3) = 1 + h^{1,1}(B_2) + {\rm rk}(G)
\eea
one finds
\bea \label{nH0-smootha}
n_H^0 = 2 + h^{1,1}(B_2) + {\rm rk}(G) - \frac{1}{2} \chi_{\rm top}(\hat Y_3). 
\eea
Together with the anomaly cancellation condition (\ref{6dgravan}) this in particular allows us to compute $n_H^c$.

Since we will make heavy use of the explicit form of $\chi_{\rm top}(\hat Y_3)$, let us briefly summarize its computation, following a general algorithm described in detail in \cite{Grassi:2000we,Grassi:2011hq}: Away from the discriminant locus, the elliptic fibration is locally a product $ \mathbb{E}_\tau \times (B_2 - \Delta)$. The product formula $\chi_{\rm top} (A \times B) = \chi_{\rm top} (A) \, \chi_{\rm top} (B)$, where $A$ and $B$ are topological spaces, implies
that $\chi_{\rm top} (\mathbb{E}_\tau \times (B_2 - \Delta)) = 0$ because  $\chi_{\rm top}(\mathbb{E}_\tau)=0$ for the generic, i.e. smooth, elliptic fiber.
Therefore  $\chi_{\rm top}(\hat Y_3)$ receives contributions only from the degenerate fibers in codimension one and two, which must be added up carefully, avoiding double counting and correcting for potential singularities of the discriminant as a divisor on $B_2$.
The result of this computation is the expression  \cite{Grassi:2000we,Grassi:2011hq}
\begin{align}\label{eq:chi}
\begin{split}
\ki (\hat Y_3)  
&= \Big( \sum_i B_i \cdot \ki (X_{P_i}) \Big) + m\, \Big(2-2g-\sum_i B_i\Big) \\
&\quad - 132 K_B^2 + m \,K_B \cdot \Sigma_1 + 2 m \,\Sigma_0 \cdot \Sigma_1 + m^2 \,\Sigma_1^2  + 3 \, C+ \sum_i \epsilon_i B_i .
\end{split}
\end{align}
Here  $X_{P_i} = \pi^{-1}(P_i)$ denotes the degenerate fiber of the resolution space $\hat Y_3$ over $P_i$, and  $B_i$ counts the number of points $P_i$ of a given type. $C$ is the number of cuspidal points $Q$ of the residual discriminant $\Sigma_0$ and the coefficients $\epsilon_i$ correct for singularities of the discriminant at $P_i$. The computation of   $X_{P_i}$, $C$ and $\epsilon_i$ is detailed in appendix \ref{app-chi}, where we also describe the generalization of (\ref{eq:chi}) to situations with several discriminant components.

\begin{table}
\scriptsize
\centering
\begin{tabular}{ccccc}
\toprule
 & $\big(\mathrm{ord}_{\Sigma_1} (f),\mathrm{ord}_{\Sigma_1} (g)\big)$ & $\mathrm{ord}_{\Sigma_1} (\Delta)$ & Equ. of monodromy cover & Gauge algebra \\\midrule
$\mathrm{I}_0$ & $(\geq 0, \geq 0)$ & 0 & --- & ---\\
$\mathrm{I}_1$ & $(0,0)$ & 1 & --- & ---  \\
$\mathrm{I}_2$ & $(0,0)$ & 2 & --- & $\mathfrak{su} (2)$  \\
$\mathrm{I}_m, \,    m\geq 3$ & $(0,0)$ & $m$ &  $\psi^2 + (9g/2f)_{z_1=0}$ & $\mathfrak{sp}(\left[\frac{m}2\right] )$ or $\mathfrak{su}(m)$\\
II & $(\geq 1, 1)$ & $2$     & --- & ---\\
III & $(1,\geq 2)$ & $3$ & --- & $\mathfrak{su}(2)$\\
IV & $(\geq 2, 2)$ & $4$ & $\psi^2 - (g/z_1^2)|_{z_1=0}$ & $\mathfrak{sp}(1)$ or $\mathfrak{su}(3)$\\
$\mathrm{I}_0^*$ & $(\geq 2, \geq 3)$ & $6$ & $ \psi^3 + (f/z_1^2)|_{z_1=0}\cdot \psi + (g/z_1^3)|_{z_1=0}$ & $\mathfrak{g}_2$ or $\mathfrak{so}(7)$ or $\mathfrak{so}(8)$\\
$\mathrm I^*_{2n-5}, \, n\geq  3$ & $(2,3)$& $2n+1$  & $\psi^2 + \frac14 (\Delta / z_1^{2n+1})(2z_1f/9g)^3|_{z_1=0}$ & $\mathfrak{so}(4n-3)$ or $\mathfrak{so}(4n-2)$\\
$\mathrm I^*_{2n-4}, \, n\geq  3$  & $(2,3)$& $2n+2$ & $\psi^2 + (\Delta/z_1^{2n+2} )(2z_1f/9g)^2|_{z_1=0}  $ & $\mathfrak{so}(4n-1)$ or $\mathfrak{so}(4n)$\\
$\mathrm{IV}^*$ & $(\geq 3,4)   $ & $8$ & $\psi^2 - (g/z_1^4)|_{z_1}=0$ & $\mathfrak{f_4}$ or $\mathfrak{e}_6$\\
$\mathrm{III}^*$ & $(3,\geq 5)$ & $9$  & --- & $\mathfrak{e}_7$\\
$\mathrm{II}^*$ & $(\geq 4, 5)$ & $10$ & --- & $\mathfrak{e}_8$\\
non-min. & $(\geq 4, \geq 6)$ & $\geq 12$ &  --- & --- \\
\bottomrule
\end{tabular}
\caption{Kodaira-Tate classification of singular fibers, monodromy covers and gauge algebras as given in \cite{Grassi:2011hq}. If the monodromy cover does not factorise, the gauge algebra is monodromy reduced as indicated.}
\label{tab:KodairaTate}
\end{table}

\section{F-theory in Presence of Non-Crepant Singularities}\label{sec:Fnoncrepant}

In this section we discuss the physics and mathematics of F-theory compactifications with non-crepant resolvable singularities in codimension two.
We begin in Section \ref{sec:Crepant} with a brief review of the standard relation between crepant resolutions and unobstructed directions in the classical Coulomb branch of the dual M-theory. This classic material is included for completeness and the expert reader can safely jump ahead.
We then explain,   in Section \ref{subsec_non-crepantb}, the general meaning of codimension-two non-crepant resolutions from the physics perspective. 
Section \ref{mathem_background} introduces the mathematical background to quantitatively analyze such singularities. In particular we introduce the notion
 of Kleinian $\mathbb Q$-factorial  terminal  hypersurface singularities and their  Milnor-Tyurina number.
 In section \ref{sec_defshyper} we establish the presence of uncharged localised hypermultiplets at such singularities, counted by the  Milnor-Tyurina number, and we explain the meaning of the topological Euler characteristic in the presence of singularities in section \ref{sec_defs}.

\subsection{Crepant Resolutions and the M-theory Coulomb Branch} \label{sec:Crepant}

Consider F-theory on $\mathbb R^{1,5} \times Y_3$ with $Y_3$ a Calabi-Yau threefold elliptically fibered over base $B_2$.
This setup is dual to M-theory on $\mathbb R^{1,4} \times Y_3$. 
More precisely if one compactifies the 6d effective action of F-theory on a circle $S_1$ of radius $R$, the resulting theory is identified with the effective action of M-theory on $Y_3$.
The radius $R$ is the inverse of 
the volume of the generic elliptic fibre of $Y_3$, $R = 1/{\rm vol}({\mathbb E}_{\tau})$, all measured in natural units.  \cite{Vafa:1996xn,Witten:1996bn}.
  The 6d F-theory effective action is recovered as ${\rm vol}({\mathbb E}_{\tau}) \rightarrow 0$. More details of this correspondence and how to recover the F-theory effective action from M-theory can be found in \cite{Intriligator:1997pq,Bonetti:2011mw,Bonetti:2013cza}.

Suppose now, as in Section \ref{sec_Ftheorynotation}, that  the elliptic fibration $Y_3$ degenerates over the vanishing locus of the discriminant $\Delta = \Sigma_1 \cup \Sigma_0$, and that 
 the generic fibre over the divisor $\Sigma_1 \subset B_2$ exhibits a singularity associated with gauge group $G$.\footnote{Generalisations to setups with several gauge branes are obvious.}
Consider the Cartan subalgebra of its Lie algebra, $\oplus_{i = 1}^{{\rm rk}(G)}    \mathfrak u(1)_i$, with associated 6d gauge potentials ${\mathbb A}^i$.
Under circle compactification to $\mathbb R^{1,4}$,
the component ${\mathbb A}_{5}^i  =: \xi^i$ of the gauge potential along $S^1$ maps to a scalar field. Together with the vector components of ${\mathbb A}^{i}$  along 
the five extended directions it forms the bosonic part of a 5d vector multiplet $(A^i, \xi^i)$.
Unlike the 6d F-theory effective action, the 5d M-theory effective action possesses a Coulomb branch, parametrized by the vacuum expectation values (VEVs) of the scalar fields $\xi^i$. 
Consider now a resolution $\hat Y_3$ of $Y_3$.
This replaces the fibral singularities with a chain of rational curves $\mathbb P^1_i$, $i=1, \ldots, {\rm rk}(G)$. Their fibration over $\Sigma_1$ is denoted by the resolution divisors $E_i$. 
Expanding the M-theory 3-form $C_3$  and the K\"ahler form $J$ as
\bea \label{C3expansion1}
C_3 = \sum_i A^i \wedge [E_i] + \ldots, \qquad \quad J = \sum_i \xi^i \, [E_i] + \ldots
\eea
gives rise to the Cartan $U(1)_i$ vector potentials $A^i$ in the 5d M-theory effective action.
The scalars $\xi^i$ are identified with the K\"ahler moduli associated with the size of the resolution curves.

By means of this identification, a resolution of the singularity in the fibre therefore corresponds to moving in the 5d Coulomb branch by allowing for a non-zero VEV $\langle \xi^i \rangle \neq 0$, whereas the singular limit corresponds to the origin of Coulomb branch $\langle \xi^i \rangle =0$ \cite{Intriligator:1997pq}. More precisely, if the resolution is crepant, the resolved elliptic fibration $\hat Y_3$ is still Calabi-Yau and supersymmetry is unbroken. 
This describes a flat direction in the Coulomb branch. 
Along such a direction in Coulomb branch, all 5d states states which are charged under the Cartan factor $u(1)_i$ acquire a mass. 
This matches with the described field theoretic perspective on F/M-theory duality as a circle reduction as follows: A massless state in 6d maps to a Kaluza-Klein (KK) zero mode in 5d together with a full tower of KK states.
The mass of the KK zero mode of Cartan charges $q_i$ is given by 
\be \label{m0}
m_0 = \sum_i q_i  \, \xi^i.
\ee
This quantity has a simple geometric meaning: The KK zero modes in M-theory arise from M2-branes wrapped on suitable curves in the fiber. 
In particular this includes states localised in codimension two, where new curves in the fiber arise from the splitting of some of the resolution $\mathbb P^1_i$.
Their
charges $q_i$ are computed via the intersection numbers of the wrapped curve with the resolution divisors $E_i$. In view of (\ref{C3expansion1}) we can therefore identify (\ref{m0}) with the volume of this curve in the resolved space $\hat Y_3$. This is true up to a sign which is explained in   \cite{Intriligator:1997pq,Aharony:1997bx}, but which plays no role for the general argument.

Since a crepant resolution corresponds to a non-trivial volume of all fibral curves, this implies that on $\hat Y_3$ also the codimension-two matter states of the 5d theory become massive. The possible inequivalent resolutions of the singularity are in 1-1 correspondence with the different Weyl chambers along the 5d Coulomb branch \cite{Intriligator:1997pq,Aharony:1997bx,Grimm:2011fx,Hayashi:2013lra,Hayashi:2014kca, Esole:2014bka, Esole:2014hya,Braun:2014kla,Lawrie:2015hia,Braun:2015hkv}. Note furthermore that in the singular F-theory limit, one recovers massless matter states in the 6d effective action.

\subsection{The Physics of Non-Crepant Resolvable Singularities in Codimension Two} \label{subsec_non-crepantb}

We can now elucidate the meaning of codimension-two singularities in F-theory which lack a crepant resolution.
As a result of what we said in the previous subsection, such singularities arise whenever there is {localized massless matter in the 6d effective action of F-theory which cannot acquire a mass in a supersymmetric way along a Coulomb branch in the dual M-theory}. 
This in particular implies that the matter state is not charged under any massless Cartan or non-Cartan abelian gauge group factor in the dual M-theory.
From the previous discussion it is clear that in such a case all flat directions along the Coulomb branch leave the matter state in question massless in M-theory.
This points to the existence of vanishing cycles in the fibre of the elliptic fibration which cannot be resolved into a non-zero volume holomorphic curve without destroying the Calabi-Yau condition, i.e. without breaking supersymmetry.

An interesting class of examples includes situations where the matter state in question carries charge only under a so-called massive $U(1)$ or, more generally, only under a discrete $\mathbb Z_k$ symmetry in the M-theory effective action.\footnote{The distinction between massless abelian gauge groups in M-theory and F-theory is important, as described in detail in the context of discrete symmetries in F-theory versus M-theory in \cite{Mayrhofer:2014haa,Mayrhofer:2014laa}.}  
Indeed, suppose the matter field is charged under a 5d gauge multiplet $(A^{\rm m},\xi^{\rm m})$ with mass $m$. Clearly, the mass term of the scalar field $\xi^{\rm m}$
obstructs the Coulomb branch and enforces $\langle \xi^{\rm m}\rangle = 0$.
Such massive vector multiplets are described in M-theory by expanding the 3-form $C_3$ along a pair of non-harmonic 2- and 3-forms $(\tw_2, \alpha_3)$ as  \cite{Grimm:2010ez,Camara:2011jg,Grimm:2011tb}
\be
C_3 = A^{\rm m} \wedge \tw_2 + c  \, \alpha_3
\ee
with the property 
\be
d \tw_2 = k \alpha_3, \qquad k \in \mathbb Z.
\ee
This relation identifies $\alpha_3$ as a torsional form, i.e. as an element of ${\rm Tor} (H^3( Y_3, \mathbb Z))$.
Dimensional reduction of the kinetic term for $C_3$ yields a St\"uckelberg coupling of the form
\be
S \simeq \int_{\mathbb R^{1,4}} (d c + k A^{\rm m}) \wedge \ast (d c + k A^{\rm m}),
\ee 
which signals the presence of a $\mathbb Z_k$ symmetry in the 5d effective action: 
The shift symmetry enjoyed by the axionic field $c$ is gauged,
\be
A^{\rm m} \rightarrow A^{\rm m} + d \chi, \qquad c \rightarrow c - k \chi
\ee
and $c$ becomes the longitudinal component of the massive vector field $A^{\rm m}$.
Note that the special case $k=1$ corresponds to a complete breaking of the $U(1)$ gauge symmetry associated with $A^{\rm m}$, whereas for $k >1$ a remnant $\mathbb Z_k$ symmetry governs the 5d (and 6d) effective action.

Now, if some massless matter state localised in codimension two carries charge only under the remnant $\mathbb Z_k$ symmetry, and is uncharged under any other massless $U(1)$ gauge symmetry, then by the above arguments the Calabi-Yau $Y_3$ will exhibit a non-crepant resolvable singularity in codimension two. 
Examples of such singularities have already appeared in the literature: For the special case $k=1$, the ${\rm I}_1$-model studied in \cite{Braun:2014nva} (and further in \cite{Martucci:2015dxa}) contains a non-crepant conifold singularity which, from a Type IIB perspective, is expected to host localised matter. Indeed, ref. \cite{Braun:2014nva} stressed the relation between the presence of a massive $U(1)$ symmetry and the occurrence of a conifold singularity in codimension-two which only admits a small non-K\"ahler resolution.
The cases $k>1$ correspond to the singular Weierstrass models \cite{Braun:2014oya} forming the Jacobian of torus-fibrations with a $k$-section: In the Weierstrass model the location of matter charged only under $\mathbb Z_k$ symmetry leads to a non-crepant singularity in codimension two.\footnote{Note that, by contrast, the associated $k$-section fibrations are smooth because these describe a different M-theory background in which the matter field in question does carry charge under a massless $U(1)$ symmetry \cite{Morrison:2014era,Mayrhofer:2014haa,Mayrhofer:2014laa} (see also \cite{Anderson:2014yva}). The models agree only in the F-theory limit, but the geometry as such is sensitive to the M-theory effective action as opposed to its F-theory uplift. Geometries associated with discrete symmetries in F-theory have been studied intensively recently, including \cite{Braun:2014oya,Morrison:2014era,Anderson:2014yva,Mayrhofer:2014haa,Mayrhofer:2014laa,Cvetic:2015moa,Cvetic:2016ner}. }

In the sequel we will find further examples non-crepant resolvable singularities in co-dimension two and establish the presence of massless matter responsible for the singularity.

\subsection{Mathematical Background} \label{mathem_background}

In this work we analyze  mostly non-smooth Calabi-Yau threefolds.
  While there are many examples of smooth Calabi-Yau varieties,  more generally, elliptically fibered Calabi-Yau varieties 
 have singularities. We have already motivated the appearance of such singularities in our context of codimension-two singular fibers.
 Independent of this, in compactifications on higher-dimensional singular spaces, new striking features appear in the associated physics, which illustrates the importance of  studying such singularities even further:  For instance \cite{Denef:2005mm} presents an example for which the crucial features of the physics are 
 captured by a singular Calabi-Yau (with terminal singularities),  but not by the  smooth birationally equivalent minimal one.  In fact, in dimension higher than three, a minimal model can be smooth while other models in the same
 minimal class are singular.  Other recent appearances of non-resolvable singularities in F-theory with rather different physics interpretation include  \cite{Garcia-Etxebarria:2015wns,LustTFects,Morrison:2016lix}.
 
 In this section we provide some mathematical background on the types of codimension-two singularities which we will study in this paper.

  \subsubsection{ $\bbQ$-Factoriality, Canonical and Terminal Singularities} \label{mathem_background1}
 
Let  $X$ be a complex algebraic (normal) variety. A {\it Weil divisor} is a formal linear combination of codimension-one subvarieties. A
 {\it Cartier divisor} on the other hand has the property that it can be locally written as the vanishing locus of a single function on $ X$.  If $X$ is smooth, then all Weil divisors are Cartier; this is true more generally if $X$ is {\it factorial}, that is if every local ring is a unique factorization domain \cite{Shafa}.
On the other hand, if $X$ is the  singular quadric in $\bbP ^3$ of equation ${ x_0 \cdot x_1- x_2^2=0 }$, the weighted projective surface $\bbP[1,1,2]$, the divisors $D_0$ and $D_1$, defined by the equations $x_0=x_2=0$ and  $x_1=x_2=0$  are Weil divisors, but not Cartier. Note that $2D_0$ and $2D_1$ are Cartier.
We say that $ X$ has   $\bbQ$-{\it factorial} singularities if for every Weil divisor $D$ there exists an integer $r$ such that $rD$ is Cartier, or equivalently that 
every  Weil divisor is $\mathbb Q$-Cartier.

Let  $X$ be a complex algebraic variety. A resolution 
of $X$ is a birational morphism $
\rho: \widetilde X \rightarrow X$ from a smooth variety $\widetilde X$ to $X$.  
This means that there  are (dense) open sets $\mathcal V \subset X$ and  $\mathcal U \subset  \widetilde X$  such that   $ \rho_{\mathcal U} : \mathcal U  \simeq  \mathcal V$. The remaining locus $  {\rm Ex} (\rho)=\widetilde X \setminus \mathcal U$ is the exceptional locus of $\rho$.  The components of codimension one in the exceptional locus are the exceptional divisors $E_i$.

 $X$ is $\bbQ$-Gorenstein if there exists some integer $r$ such that $rK_X$ is a line bundle (that is $K_X$ is $\bbQ$-Cartier). When $K_X$  is Cartier $X$ is said to be {\it Gorenstein}. In particular a  Calabi-Yau variety is Gorenstein.
 For a $\bbQ$-Gorenstein variety $X$ and its resolution $\tilde X$ we can compare 
 the bundles  $rK_{\widetilde X}$ and $rK_X$,
  \bea
rK_{\widetilde X} = \rho^*rK_X + \sum_i a_i  r E_i \,.
\eea
The  $a_i$ are called the {\it discrepancies}. Such a resolution always exists \cite{Hironaka, BierstoneMilman} and  it is easy to see that  the discrepancies are independent of the choice of the resolution.

If  $a_i \geq 0 \ \forall i$, $X$ is said to have  {\it{at worst canonical }} singularities. If $a_i > 0$ for all $i$, $X$ is said to have {\it at worst terminal} singularities. Reid showed that if $X$ has at worst canonical singularities, then $H^0(mrK_X)=H^0(mrK_{\widetilde X}), \ \forall \  m \in \bbN$.

 Note that a smooth variety has at worst terminal singularities.
If $X$ is a surface it can be shown that  $X$ has at worst terminal singularities if and only if it  is smooth. The canonical  surface singularities are the A-D-E singularities (rational double points). These are also the Gorenstein surface singularities, see for example \cite{HaconKovacs, KollarMori}. In particular   if  $\dim (X)=2$ and $X$ is a  Weierstrass model with singularities,  then  its singularities are canonical.

  If for a particular resolution  $a_i=0 \ \forall i$, the resolution is called {\it crepant}, and  we refer to these singularities  as crepant resolvable. A crepant resolution of a singular Calabi-Yau variety hence remains Calabi-Yau. 
For example  an elliptically fibered  threefold $X$ given by a Weierstrass model which is  equisingular along a smooth curve has canonical singularities  and a crepant resolution.  
More generally a Weierstrass model has canonical singularities in codimension one. 
 This means that the non-crepant resolvable singularities of such a model must be due to enhancements in codimension two or higher.

A morphism $ \varphi:  \widetilde X  \to X$ where $\widetilde X$ is smooth and all the components of the exceptional loci  have codimension greater than one is called  a {\it small resolution}. In this case $\widetilde X$ and $X$ are   {\it isomorphic in codimension one}. For example,
the  nodal quintic threefold $X \subset \bbP ^3$ of equation $ x_0g_0 + x_1 g_1=0$  with general polynomials $g_0$ and $g_1$ has 16  nodal isolated singularities and a small resolution
$ \phi:  \widetilde X  \to X$, obtained by blowing up the plane $x_0=g_1=0$; the exceptional loci consist of $16$ disjoint ${\bbP}^1$s.  
 Because small resolutions of singularities preserve the Calabi-Yau condition, we also refer to the corresponding singularities as  resolvable by a crepant resolution.  
 Alternatively, we could  resolve $X$ by a big resolution with exceptional divisors. However,  the  nodal singularities are terminal in the sense that the appearing 
  exceptional divisors have positive discrepancy.

 A small algebraic resolution of the nodal quintic threefold above is possible because the singularities are not $\bbQ$-factorial, as we see in the following.

\subsubsection{  $\bbQ$-factorial and  analytic $\bbQ$-factorial singularities} \label{mathem_background2}

In the class of birationally equivalent elliptic fibrations of Calabi-Yau threefolds there is a model $ X$ with $K_{X} \simeq \mathcal O_{X}$ where $X$ has terminal singularities, but $X$ is not necessarily smooth \cite{Grassi1991, GrassiEqui}. 
  The model $X$ has  however  $\bbQ$-{\it factorial} singularities, namely every  Weil divisor is also $\mathbb Q$-Cartier.  
      Kawamata \cite{Kawamata1988}   showed that if the singularities are at worst canonical\footnote{Kawamata proves the result assuming that the singularities are rational; canonical and terminal singularities are rational.} then the quotient  of the Weil divisors by the Cartier divisors 
has finite dimension,  and its rank is denoted by $\sigma (X)$. $X$ is $\bbQ$-factorial if and only  this group is torsion, that is if  $\sigma (X)=0$. Kawamata shows that if $X$ is not $\bbQ$-factorial, there exists a small projective (K\"ahler) birational morphism $\phi: X_1 \to X$, where $X_1$ is $\bbQ$-factorial.
  For example the  nodal quintic threefold $X: \   x_0g_0 + x_1 g_1=0$ is not $\bbQ$-factorial, as the divisor  $x_0=g_1=0$ must be defined by two equations; the small birational morphism $\phi$ provides a small projective  resolution.
 When the isolated singularity is {\it toric}, then there is a nice criterion: the singularity is $\bbQ$-factorial if and only if the maximal cone corresponding to the toric singular point is {\it simplicial}. If the cone is not simplicial, the small resolution is achieved by a simplicial subdivision of the cone. 
 
 In many instances, a Calabi-Yau threefold is $\bbQ$-factorial,  but after an analytic change of coordinates the local equation is \bea
f(z,x_1,x_2,x_3) = z^2 + x_1^2 + x_2^2 + x_3^2.
\eea 
Such behaviour characterizes the singularities in the Weierstrass model of \cite{Braun:2014nva}, studied in Section \ref{sec_conifold},  as well as the singular Jacobians of \cite{Braun:2014oya}. Both singularities correspond to fiber enhancements ${\rm I}_1 \times {\rm I}_1 \rightarrow {\rm I}_2$. Another example of this behaviour which we will study is a particular enhancement of type ${\rm III} \times {\rm I}_1 \rightarrow {\rm I}_0^*$ in Section \ref{sec_typeIIImodel}.
In these examples there exist two local independent analytic, but not algebraic, Weil divisors which are not $\mathbb Q$-Cartier.
This motivates the following definition: 
$(\mathcal U, p)$ is  {\it locally analytically $\mathbb Q$-factorial} if every analytic Weil divisor in a neighborhood of $p$ is $\mathbb Q$-Cartier. 
 
More generally, an  important class of isolated hypersurface singularities  are the  $A_{a-1}$ Kleinian singularities:  In  a neighborhood  of the singular point $P$, $(\mathcal U, p)$ is (analytically) the zero-locus of
\bea \label{A_a-1sing}
f(z,x_1,x_2,x_3) = z^a + x_1^2 + x_2^2 + x_3^2 \qquad {\rm with} \qquad a \geq 2 \in \mathbb N \,.
\eea 
These are terminal (and non-canonical) singularities \cite[Th.~1.1]{ReidCanonical}. A local, possibly non-projective (non-K\"ahler) small resolution is possible if and only if $a$ is even, \cite[Cor.~1.6]{ReidCanonical}, \cite{Atiyah, Brieskorn}.
In particular if $a$ is odd, then no crepant or small resolution is possible and  these singularities are also $\bbQ$-factorial and analytically $\mathbb Q$-factorial \cite{Flenner}. 
The isolated singularities in this paper happen to be Kleinian: These are the  Kodaira fiber ${\rm II} \times {\rm I}_1 \rightarrow {\rm III}$ and ${\rm II} \times {\rm I}_2 \rightarrow$ IV enhancements in the models of Section \ref{typeIImodel} as well as certain types of Kodaira fiber ${\rm III} \times {\rm I}_1  \rightarrow {\rm I}_0^\ast$ enhancements in Section \ref{sec_nontrivialgaugegroup}. 

If $a$ is even, or more generally for other hypersurface equations, a careful global analysis is needed to determine if a  projective  (K\"ahler) small resolution exists, see for example \cite{LinUFD}.

Finally let us note an important point:
Given a three-dimensional Calabi-Yau Weierstrass model, it is always possible to resolve the singularities in codimension one in a crepant way. However, there may remain $\mathbb Q$-factorial terminal singularities in codimension two.
These are always analytic hypersurface singularities, see for example \cite{KollarMori}.
Technically, this means that a Weierstrass model can always be resolved into a terminal $\mathbb Q$-factorial model. 
The Kleinian singularities we are studying in this paper are a special type of these $\mathbb Q$-factorial terminal hypersurface singularities.

\subsubsection{Milnor and Tyurina Numbers, and Versal Deformations}\label{mathem_background3}

An important concept for us is the characterization of hypersurface singularities via their Milnor and Tyurina numbers.
Let $\mathcal U$ be a neighborhood of an isolated hypersurface singularity $P$, that is ${\mathcal U}=  f^{-1}(0)$, where $f: \mathbb C^{n+1} \rightarrow \mathbb C$, and consider a local smoothing $\mathcal U_t=  f^{-1}(t)$. 
Let $D_\epsilon$ denote a ball of radius $\epsilon$ centered at $0 \in \mathbb C^{n+1}$.
Milnor showed that  for $\epsilon >0$ small enough,  $B_\epsilon = {\mathcal U}^t     \cap D_\epsilon$   is homologically a bouquet of $n$-spheres, where   $B_\epsilon$ is called the {\it  Milnor fiber }of $P$. The {\it Milnor number} characterizing this singularity is
\bea
m_P = b_n(B_\epsilon),
\eea
the $n$th Betti number of the ordinary simplicial homology.
Equivalently \cite{Looijenga2013},
\begin{align*}
m_P = \dim_{\mathbb C} \mathcal A_f = \dim_{\mathbb C} \left(\mathbb C\{ x_1, \ldots, x_{n+1} \}/ \left\langle  \frac{\partial f}{\partial x_1}, \ldots   \frac{\partial f}{\partial x_{n+1}}  \right\rangle \right).
\end{align*}
 For example, an  $A_{a-1}$ Kleinian  singularity  as in \eqref{A_a-1sing} has Milnor number  $m_P = a-1$, since 
\begin{align*}
\mathcal A_f = \mathbb C\{ z,x_1, \ldots, x_{3} \} / \langle z^{a-1},x_1,x_2,x_3 \rangle = \langle 1,z,z^2,\ldots,z^{a-2} \rangle
\end{align*}
and the number of generators of $\mathcal A_f$ is $a-1$ .

A related concept, from the algebraic point of view, is the {\it Tyurina number}  $\tau_P$, which counts the dimension of the space of versal deformations of the hypersurface singularity at $P$ in $\mathcal U$.  
The Tyurina number is computed algebraically as
\bea
\tau_P = \dim_{\mathbb C} \mathcal B_f = \dim_{\mathbb C} \left(\mathbb C\{ x_1, \ldots, x_{n+1} \}/ \left\langle f, \frac{\partial f}{\partial x_1}, \ldots   \frac{\partial f}{\partial x_{n+1}}  \right\rangle\right).
\eea
In general $\tau_P \leq \mu_P$. Saito showed that $\tau_P = \mu_P$ if and only if    $P$ is a weighted hypersurface singularity, that is if
there exist weights $(d_1, \ldots d_{n+1})$ and $d$ such that 
 \be
 f(\lambda^{d_1} x_1, \ldots, \lambda^{d_{n+1}} x_{n+1} ) = \lambda^d f(x_1, \ldots x_{n+1})
 \ee
  for all $\lambda \in \mathbb C$ \cite{Looijenga2013}. A generalization of this result for complete intersections is proven by Greuel \cite{Greuel1980}. The Tyurina and Milnor number can be computed by SINGULAR \cite{GreuelLossenShustin2007}  and Maple \cite{RossiTerracini}.

In particular, the $A_{a-1}$ Kleinian singularities  of  \eqref{A_a-1sing} are weighted hypersurface singularities, and hence the Tyurina and Milnor numbers agree,
\be \label{mp=tp}
\mu_P = \tau_P = a-1     \qquad {\rm for} \qquad \eqref{A_a-1sing} \,.
\ee

\subsection{ Hypermultiplets in Presence of $\mathbb Q$-Factorial Terminal Singularities } \label{sec_defshyper}
      
 The presence of singularities makes the computation of the spectrum of massless moduli fields of a string compactification more involved.
As reviewed in Section \ref{sec_Ftheorynotation}, in F-theory on a smooth Calabi-Yau threefold $X$, the number of tensor and uncharged hypermultiplets is related to the 
dimensions of the space of K\"ahler deformations and of complex structure deformations. 
We now describe the situation in the presence of isolated hypersurface singularities and describe methods to compute in particular the hypermultiplet spectrum in the presence of 
$\bbQ$-factorial  Kleinian terminal singularities of type $A_{a-1}$ with either $a$ odd or $a=2$, as these are the type of singularities which occur in the examples we consider here .The methods developed in  \cite{ArrasGrassiWeigandM} are, however, more general, and in particular hold for rational homology manifolds.

\subsubsection{K\"ahler deformations of Singular Threefolds} \label{sek_defs}

For F-theory compactified on a  smooth Calabi-Yau threefold $X$,  with  zero  Mordell-Weil rank, 
 the number of tensor multiplets $n_T$  is given by
\bea \label{nT1}
n_T+2 + \operatorname{rk} (\mathfrak g)=\kd(X) \,,
\eea
with $\kd(X)$ the dimension  of the space of K\"ahler deformations.
  This formula continues to hold 
in the presence of isolated singularities on a threefold $X$.
When  $X$ is smooth then 
$ \kd(X)= h^{1,1}(X)=b_2(X), $
   where $b_2(X)$ is the second Betti number of the ordinary (simplicial) homology. It turns out that if $X$ is a  Calabi-Yau with  terminal singularities, and under more general assumptions spelled out in the companion paper \cite{ArrasGrassiWeigandM},  the equalities
 \begin{align}\label{KdefY3}
   \kd(X)= b_2(X)
\end{align}
   and 
\bea
n_T+2 + \operatorname{rk} (\mathfrak g)=b_2(X)
\eea    
continue to hold.
  
\subsubsection{ Hypermultiplets and Complex Deformations} \label{sec_defs}

More subtle is the number $n_H^{0}$ of uncharged hypermultiplets.
 On a smooth Calabi-Yau threefold $X$, this is related to the  dimension of the Kuranishi space of complex structure deformations $\cd(X)$ via 
\bea  \label{unchm}
 n_H^{0} =  1 + \cd(X) \,.
 \eea
When
  $X$ is smooth, it is furthermore  a classic result  that $\cd(X)= h^{2,1}(X)= \frac{1}{2} b_3(X) -1$.

Now, in the presence of isolated singularities on a threefold $X$, the relation (\ref{unchm}) is also still valid as each complex structure deformation corresponds to a massless modulus of the metric.  
However, in the presence of singularities one cannot use the formula   $\cd(X)= h^{2,1}(X)= \frac{1}{2} b_3(X) -1$ to compute the number of complex structure deformations. The reason for this is explained in more detail in \cite{ArrasGrassiWeigandM}. We show now how to calculate  $\cd(X)$ when $X$ is  a singular Calabi-Yau with $\bbQ$-factorial terminal hypersurfaces singularities. Note that  the resolutions  of the general singularities (in codimension two) of Weierstrass models have indeed isolated hypersurface singularities.
The proofs  are presented in the companion paper \cite{ArrasGrassiWeigandM}.

Results of Namikawa and Steenbrink   \cite{NamikawaSteenbrink} imply that if $X$ is a $\bbQ$-factorial Calabi-Yau threefold with isolated terminal hypersurface singularities,   $X$ admits a smoothing to a smooth Calabi-Yau $X^t$.
 The dimension of the complex deformation of $X$ is  then given by the dimension  of the complex deformation space of $X^t$.   Since $\cd(X^t) = h^{2,1}(X^t) = \frac{1}{2} b_3(X^t) - h^{3,0}(X)$, we then have the relation
\bea\label{XXt}
\cd (X)= \cd (X^t) =\frac{1}{2} b_3(X^t)-1.
\eea 

Namikawa and Steenbrink show  more generally that  if $X^t$ is a smooth deformation of   a (normal projective) threefold $X$  with isolated hypersurface singularities and 
 $h^2(X, \mathcal O_X)=0$, then
\bea\label{b3b3t}
b_3(X^t) = b_3(X) + \sum_P m_P -  \sigma(X) \,.
\eea 
Here  $\sigma(X) $ is the rank of the quotient of the Weil divisors by the Cartier divisors,  and $m_P$ is the Milnor number of the singular point $P$, both defined in Section \ref{mathem_background3}. 
Combining equations (\ref{XXt}) and (\ref{b3b3t}) we obtain:
 \bea
\cd(X) = \frac{1}{2} \big( b_3(X) + \sum_P m_P -  \sigma(X) \big)  - 1.
\eea 
Under the same hypothesis Namikawa and Steenbrink show also that   \cite{NamikawaSteenbrink}  that $ \sigma(X) = b_4(X) - b_2(X)$; in fact, Poincar\'e duality does not necessarily hold. For example, we can use the above formula to calculate $\cd(X)$ for Calabi-Yau varieties with conifold singularities which are not  $\bbQ$-factorial; recall that $X$ is $\bbQ$-factorial if and only if $ \sigma(X) =0$.

The important point for applications in this paper is that if $X$ is a Calabi-Yau variety with $\bbQ$-factorial terminal singularities, then
 \begin{align}\label{CdefY3a}
\cd(X) = \frac{1}{2} \big(b_3(X) + \sum_P m_P \big)   - 1 \,.
\end{align}
Note that only the sum $\frac{1}{2} b_3(X) + \frac{1}{2}\sum_P m_P$ is guaranteed to be integer, whereas each individual term may fail to be so because the Hodge decomposition and  Hodge duality might not hold; these points are discussed further in  \cite{ArrasGrassiWeigandM}. 
It follows that  the number of uncharged hypermultiplets  of equation (\ref{unchm}) is given by
\be
\ba
 n_H^{0} &=  1 + \cd(X)\\
& =   \frac{1}{2} ( b_3(X) + \sum_P m_P ).
\ea
\ee

 \subsubsection{Uncharged Localised Hypermultiplets }\label{UnchLocHyper}
 
In Section \ref{subsec_non-crepantb} we had argued that in the presence of  $\bbQ$-factorial  terminal (or more generally non-crepant resolvable) codimension-two singularities we expect localised massless uncharged matter. This implies  a split of the total number of uncharged hypermultiplets into localised versus non-localised uncharged hypermultiplets of multiplicity $n_{H, l}^0$ and $n_{H,n-l}^0$:
\bea \label{split}
n_H^0 =  n_{H,n-l}^0  + n_{H, l}^0 \,.
\eea
 The uncharged localised hypermultiplets are to be interpreted as part of the Kuranishi space, i.e. the space of complex structure deformations of the singular space $X$.\footnote{We count the universal hypermultiplet as part of the $n_{H,n-l}^0$ non-localised hypermultiplets.}
 The  split (\ref{split}) implies a natural decomposition of the Kuranishi space of $X$ into two spaces $K_{n-l}$ and $K_{l}$ \cite{ArrasGrassiWeigandM}. The space $K_{l}$ is the space of complex structure deformations of $X$ which deform the isolated singularities, by changing their singularity into a milder singularity type (or completely smoothening them out). These are precisely the versal deformations, and the dimension of this space is counted by the Tyurina number.
In fact,  in our hypothesis  $m_P= \tau_P$, see (\ref{mp=tp}).
 The remaining deformations of $X$ are deformations which do not change the location or form of the isolated singularities. 

This suggests identifying the  localised uncharged hypermultiplets with the metric moduli counted by the versal deformations such that
\bea
n_{H, l}^0 = \sum_P \tau_P= \sum_P  m_P,
\eea
while the non-localised uncharged hypermultiplets are due to the remaining deformations of $X$.
We therefore find that\footnote{Note also here  that $\frac{1}{2} ( b_3(X) - \sum_P m_P)$ is an integer.} 
\bea
n_H^0 &=&  \frac{1}{2} \big( b_3(X) - \sum_P m_P\big) +   \sum_P m_P \\
&=&  n_{H,n-l}^0   + n_{H, l}^0.
\eea

The identification of the versal deformation moduli with localised hypermultiplets is indeed very natural given the general relation between deformations and Higgsings: Physically, the deformation of a singularity corresponds to a process where a massless hypermultiplet acquires a vacuum expectation value such that the singularity arises at the origin of a Higgs branch. For singularities which allow a crepant resolution the localised hypermultiplets are charged and the Higgsing necessarily breaks part of the gauge group. A classic example is the deformation of a codimension-two resolvable conifold singularity in the fiber of an F-theory elliptic fibration, corresponding to a Higgsing of a $U(1)$ gauge group under which the localised states are charged \cite{Braun:2011zm,Krause:2012yh,Intriligator:2012ue}. The novelty in absence of a crepant resolution is that the localised states are uncharged, at least under any massless gauge group.

\subsection{The Euler Characteristic of Singular Threefolds} \label{sec_defs}

In our examples $X$ is a Calabi-Yau variety with $\bbQ$-factorial  Kleinian terminal singularities of type $A_{a-1}$ with either $a$ odd or $a=2$.  
It is a consequence of a result in \cite{ArrasGrassiWeigandM}  that in both cases

\begin{align} \label{chitopAll}
\chi_{\rm top}(X) =2 + 2 b_2(X)- b_3(X),
\end{align}
as in the smooth case.
 It is important to stress that the arguments needed are more general than for Kleinian singularities, but different for $a$ odd (a rational homology manifold) and $a=2$, and the final statement is the same.
By combining  equations (\ref{chitopAll}),  (\ref{CdefY3a}) and (\ref{KdefY3})
 we find
 \begin{align}\label{eq:cd}
 \tfrac12 \, \ki (X) = \kd (X) - \cd (X) + \tfrac12 \, \sum_P m_P.
 \end{align}
In particular,
\be
\ba
 n_H^{0} &=  1 + \cd(X)\\
& =   \frac{1}{2} ( b_3(X) + \sum_P m_P ) \\
& = 1 + \kd (Y_3) - \tfrac12 \, \ki (Y_3) + \tfrac12 \, \sum_P m_P,
\ea
\ee
and 
\bea \label{loc-nonlocHyerps}
n_H^0 &=&  \left(1 + \kd (X) - \tfrac12 \, \ki (X) - \tfrac12 \, \sum_P m_P\right)   +   \sum_P m_P \\
&=&  n_{H,n-l}^0   + n_{H, l}^0. 
\eea
These expressions will be successfully be applied in a number of examples in the remainder of this paper and we will verify that  the resulting spectrum is free of gravitational anomalies. 
In particular, the methods of \cite{Grassi:2000we} to compute the Euler characteristic of the elliptic threefold, reviewed in appendix \ref{app-chi}, are still valid in the presence of singularities. 
The crucial point, however, is that the topological Euler characteristic of the fibers entering (\ref{eq:chi}) must be evaluated for the singular or  partially resolved models.  
We now turn to explaining this procedure in more detail.

\section{Terminal Singularities in Models With Trivial Gauge Group}\label{sec:modelswithoutgauge}

In this section we exemplify the appearance of uncharged, localised matter at non-crepant singularities in two models with trivial gauge group. The first model is perturbative with a terminal singularity of conifold type at an ${\rm I}_1 \rightarrow {\rm I}_2$ enhancement locus, while the second class of models is inherently non-perturbative in nature due to a terminal singularity at a II $\rightarrow$ III locus.

\subsection{The $\text I_1$ Conifold Model} \label{sec_conifold}

As our first example of a non-resolvable model, we consider a non-generic elliptic fibration with only ${\rm I}_1$ singularities in codimension one.
This model has been discussed before in \cite{Grassi:2000we,Braun:2014nva,Martucci:2015dxa}
and is most efficiently described as a specialization of a Tate model\begin{align}\label{eq:TateForm}
y^2 + a_1\, x y z + a_3 \, y z^3  = x^3 + a_2 \, x^2 z^2  + a_4\, x z^4  + a_6 \, z^6,
\end{align}
where the $a_n$ are sections of $\mathcal O (-n K_B)$.
The parameters of the Tate form and the Weierstrass form are related via
\begin{align}\label{fgfrom Tate}
f = -\tfrac1{48} (b_2^2 - 24 \,b_4), \quad g = -\tfrac1{864} (-b_2^3 + 36\, b_2 b_4 - 216\, b_6),
\end{align}
where the $b_n$ are sections of $\mathcal O (-n K_B)$ and take the form
\begin{align}
b_2 = a_1^2 + 4 a_2, \quad b_4 = a_1 a_3 + 2 a_4,\quad  b_6 = a_3^2 + 4 a_6.
\end{align}
Let us now choose the vanishing orders
\bea
{\rm ord}(a_1,a_2,a_3,a_4,a_6)|_{z_1 = 0} =: (k_1,k_2,k_3,k_4,k_6) =   (0,0,1,1,1)
\eea
along a divisor $\Sigma_1: z_1 = 0$ on the base $B_2$ by setting $a_i = \tilde a_i z_1^{k_i}$ with $\tilde a_i$ generic.
As a result, the discriminant $\Delta$ splits into two components $\Sigma_1$ and $\Sigma_0$,
\begin{align*}
\Delta = \tfrac1{16} \,z_1\, \Big( \tilde a_{6}\, (\tilde a_1^2+4\,\tilde a_2)^3 + z_1\cdot (\ldots) \Big)
\end{align*}
and in terms of the associated Weierstrass model the vanishing orders along generic points of the ${\rm I}_1$-locus $\Sigma_1$ are 
\bea
{\rm ord}(f,g,\Delta)|_{z_1=0} = (0,0,1).
\eea
There are two types of codimension-two enhancement points from the intersection $\Sigma_1 \cap \Sigma_0$,
\bea
 P_1:&  \{z_1=0\} \cap \{ \tilde a_1^2+4\,\tilde a_2=0 \}     \quad &{\text I}_1 \rightarrow {\rm II}  \\
 P_2:& \{z_1=0\}\cap \{\tilde a_{6}=0\}  \quad    &{\text I}_1 \rightarrow {\text I_2}.
\eea
At $P_1$ the singular fiber takes the form of a  singular cuspidal curve with $\ki (X_{P_1}) = 2$ even though the threefold $Y_3$ remains smooth.
At $P_2$ the fiber develops an $\text I_2$ singularity, corresponding to a singularity of $Y_3$. This singularity admits no small resolution (see~\cite{Grassi:2000we,Braun:2014nva}). In the language of Section (\ref{mathem_background}), it is in fact $\mathbb Q$-factorial terminal (but not analytically $\mathbb Q$-factorial). 
Indeed the singularity is only locally of conifold form  
\bea
z^2+x_1^2+x_2^2+x_3^2=0
\eea
with higher order terms in $z$ obstructing a small resolution \cite{Braun:2014nva}. 
From a physical perspective the absence of a small, crepant resolution is owed to the fact that the gauge group in the M-theory compactification is trivial and hence no Coulomb branch is available to render the expected massless states localised at $P_2$ massive.

The Milnor number of this $A_1$ type singularity is $m_{P_2}=1$.
In order to compute the number of complex structure deformations $\cd$ of the singular threefold $Y_3$ we  follow the formalism developed in Section \ref{sec_defs} and in particular apply (\ref{eq:cd}), where $\tfrac12 \, \sum_P m_P = \tfrac12 \, \sum_{P_2} m_{P_2}  = \frac{1}{2} \, (-6 K_B - [z_1]) \cdot_{B_2} [z_1]$, which is just $\frac{1}{2}$ times the number of points $P_1$.

For concreteness let us now take $B_2=\mathbb P^2$ with homogenous coordinates $[z_0 : z_1 : z_2]$ and identify
$\Sigma_1: z_1 =0$. This is a rational curve and hence $g(\Sigma_1) = 0$.\footnote{In particular there are no bulk matter states propagating along $\Sigma_1$. }
 There are then 6 points of type $P_1$ and 17 points of type $P_2$. Interestingly,  $\ki (Y_3)$ must now be odd. This reflects the fact that on the singular space $Y_3$ ordinary homology does no longer enjoy Poincar\'e duality, in agreement with the fact that $Y_3$ is only $\mathbb Q$-factorial, but not analytically $\mathbb Q$-factorial.  
 
 This is indeed confirmed by explicit computation of $\ki (Y_3)$ via (\ref{eq:chi}) as follows:
Since the fiber $X_{P_2}$ over $P_2$ is not resolved, we must take $\ki (X_{P_2}) = 1$, corresponding to the value of the singular original fiber. 
The intersection multiplicities of $f$ and $g$ at both points are given by $\mu_{P_1}(f,g) = 2,\, \mu_{P_2}(f,g) = 0$ \cite{Grassi:2000we}, and the parameters $\epsilon_i$ correcting for singularities of $\Delta$ at $P_i$ are given by $\epsilon_1 = -1 = \epsilon_2$ (see Table 4 in \cite{Grassi:2000we}).
All in all, this leads to $\ki (Y_3) = -523$.
Taking into account that $h^{1,1}(Y_3) =2$ (since the gauge group is trivial), 
 our expression (\ref{eq:cd}) for the number of complex structure deformations yields
\begin{align*}
\cd (Y_3) = 2+\tfrac12 \cdot 523 + \tfrac12 \cdot 17 \cdot 1 = 272.
\end{align*}
Since the gauge group is trivial, there are no charged hypermultiplets at all, and we find
\bea
n_H = n_H^0 = 1 + \cd (Y_3) = 273
\eea
in agreement with condition (\ref{6dgravan}) for cancellation of gravitational anomalies.  
According to (\ref{loc-nonlocHyerps}), the hypermultiplets from  $\cd (Y_3)$ split into $17 \times 1$ localised uncharged hypermultiplets, while the remaining ones are unlocalized states. 
This fits perfectly with the type IIB orientifold limit of the Tate model, as described in detail in \cite{Braun:2014nva}: On the Calabi-Yau twofold which is the Type IIB double cover of $B_2$, $\Sigma_1$ uplifts to two divisors $D_1 \cup D_1'$ exchanged by the orientifold action, while the 7-brane along $\Sigma_0$ uplifts to the O7-plane together with another 7-brane on an invariant divisor $D_0$ of Whitney type. The 17 points $P_2$ correspond to the intersection points between $D_1$ and $D_0$ (which are identified with $D_1' \cap D_0$ by the orientifold involution), each of which gives rise to one massless hypermultiplet from strings streched between both branes.\footnote{The points $P_1$ uplift to the intersection points between $D_1$ and $D_1'$ on top of the orientifold plane; here no additional massless matter states reside as the 7-7' string zero modes are projected out by the orientifold action.} Since the $U(1)$ gauge symmetry from $D_1$ and $D_1'$ is massive by a St\"uckelberg mechanism, this matter appears as uncharged in F/M-theory, but still localised.

\subsection{The Type II Model} \label{typeIImodel}

In this section we consider two non-crepant resolvable Weierstrass models with trivial gauge group which do not allow for a perturbative Type IIB orientifold limit. 
Over the divisor $\Sigma_1$ we engineer a type II (cuspidal) Kodaira fiber, which shall    
 enhance in codimension-two  to type III (model 1) or to type IV (model 2). This is achieved by the following vanishing orders\footnote{More generally, we can consider $f= z_1^n \, f_0$, $g = z_1 \, g_0$. The models with $n > 2$ are similar to $n=2$ and are discussed in appendix \ref{[n1]models}.} in the Weierstrass model:
\begin{align*}
{\rm model \,  1:\, } &\qquad    f = z_1 f_0, \,\quad  g= z_1 g_0  \qquad   \rightarrow &\Delta = z_1^2 \cdot \big( 27g_0^2 + 4 z_1 f_0^3 \big) \\
{\rm model \, 2: \, } &\qquad    f = z_1^2 f_0, \,\quad  g= z_1 g_0 \qquad  \rightarrow  &\Delta = z_1^2 \cdot \big( 27g_0^2 + 4 z_1^4 f_0^3 \big)
\end{align*}
There is one type of intersection points $P_1$ of $\Sigma_1$ and $\Sigma_0$ at $z_1= g_0 = 0$:
In model 1, the vanishing orders of $(f,g,\Delta)|_{P_1}$ are $(1,2,3)$, corresponding to type III, while in model 2, $(f,g,\Delta)|_{P_1} = (2,2,4)$, indicating an enhancement to type IV.
Since the gauge group is trivial, we expect the isolated singularities at $P_1$ not to allow for a crepant resolution. Indeed, in both models the singularity at $P_1$ can be brought into the form of a hypersurface singularity
\bea
z^3+x_1^2+x_2^2+x_3^2=0 .
\eea
To see this, note that locally near the singularity at $x=y=z_1=g_0=0$ we can set $f_0=1$ and $z=1$ and rewrite the Weierstrass polynomial as $P_W = -y^2   + x^ 3 + \frac{1}{4}[z_1 + ( z_1^k \, x + g_0)]^2    -   \frac{1}{4}[z_1 - (z_1^k \, x + g_0)]^2   $ with $k=0$ and $k=1$ for model 1 and 2, respectively.
As described after equ. (\ref{A_a-1sing}), such a singularity is analytically $\mathbb Q$-factorial terminal, with Milnor number $\mu_{P_1} = 2$. 
We can therefore use (\ref{loc-nonlocHyerps}) to determine the number of localised and unlocalised neutral hypermultiplets.

To this end we first evaluate $\chi_{\rm top}(Y_3)$ via (\ref{eq:chi}) with the help of the data summarized in table \ref{tab:oneBraneResultsnonResNogauge}.
In model 1, the residual determinant $\Sigma_0$ is smooth at $P_1$ and hence the parameter $\epsilon_1$  defined in more detail in appendix \ref{sec:howToComputeEpsilon}, especially equ. (\ref{eq:epsilonFormulae}), is $\epsilon_1 = -1$. In model 2, $\Sigma_0$ at $P_1$ is locally of the form $x^2 + y^4 = 0$ so that $\epsilon_1=2$.
The intersection multiplicity $\mu (f,g)$ vanishes since $f_0$ and $g_0$ are generic at $P_1$. 
The topological Euler characteristic of the fiber over the enhancement points is $\ki (X_{P_1}) = 2$, corresponding to the value of $\ki$ of the type II fiber, since both models are not resolvable. 

At this stage we restrict ourselves, for concreteness, to $B_2 = \mathbb P^2$ and take again $\Sigma_1: z_1=0$ with $z_1$ one of the homogeneous coordinates $[z_0 : z_1 : z_2]$. This implies that $g(\Sigma_1)=0$ and the number of points of type $P_1$ is $B_1 = 17$.
For both models\footnote{The difference in $\epsilon_1$ is compensated by a different number $C$ of cuspidal points $Q$ appearing in (\ref{eq:epsilonFormulae}) since $\mu_f$ differs in both cases.}  this leads to $\ki (Y_3) = -506$.
Thus, the number (\ref{eq:cd}) of complex structure deformations is 
\begin{align*}
\cd (Y_3) = 2+\tfrac12 \cdot 506 + \tfrac12 \cdot 17 \cdot 2 = 272.
\end{align*}
As always there is also the universal hypermultiplet so that $n_H = n_H^0 = 273$, as required by the gravitational anomaly condition. 
This time, since $m_{P_1}=2$, the number of localised uncharged hypermultiplets per terminally singular point is $2$ - a statement which is of course independent of the choice of base space. 
The 273 hypermultiplets thus split into $n^0_{H,l} = \sum_{P_1} m_{P_1} =  17 \cdot 2 = 34$ uncharged localised and $n^0_{H,n-l} =  1 + 272-34 = 1 + 238$ non-localised ones. All results of this section are summarized in Table~\ref{tab:oneBraneResultsnonResNogauge}.

\begin{table}
\centering
\begin{tabular}{lrr}
\toprule
$(\mu_f ,\mu_g)$    & $(1,1)$ & \hspace{3em}$(2,1)$\\
$m$                 & $2$     & $2$     \\\midrule
Enhancements        & II $\to$ III     & II $\to$ IV\\
$h^{1,1} (\tilde Y_3)$ & $2$  & $2$     \\
$B_1$               & $17$    & $17$    \\
$\ki(X_{P_1})$      & $2$     & $2$     \\
$\epsilon_1$        & $-1$    & $2$    \\\midrule
$\ki (Y_3)$         & $-506$  & $-506$  \\
$a$                 & 3       & 3       \\
$m_P$               & $2$     & $2$     \\\midrule
non-localised hypers $n^0_{H,n-l}$ & $239$& $239$ \\
localised hypers $n^0_{H,l}$ & $34$& $34$    \\\midrule
$273-(n^0_{H,n-l}+n^0_{H,l})$ & 0   & 0    \\\bottomrule
\end{tabular}
\caption{Non-resolvable models with trivial gauge group. The parameter $a$ characterises the form of the terminal codimension-two hypersurface singularity at $P_1$ via $z^a+ x_1^2+x_2^2+x_3^2=0$. $m_P$ denotes the corresponding Milnor number.}
\label{tab:oneBraneResultsnonResNogauge}
\end{table}

\section{Terminal Singularities in Presence of Non-Trivial Gauge Group} \label{sec_nontrivialgaugegroup}

In this section we present a family of models with $\mathbb Q$-factorial terminal codimension-two singularities in presence of a non-trivial gauge group. 
According to our general logic the $\mathbb Q$-factorial terminal singularities should host both charged and in addition uncharged localised matter.
This expectation is indeed confirmed by our explicit analysis.

\subsection{A Family of Type III Models With $\mathbb Q$-Factorial Terminal Singularities} \label{sec_typeIIImodel}
In the setup of interest  the discriminant exhibits a (non-perturbative) type III singularity along a divisor $\Sigma_1: z_1 = 0$, corresponding to gauge group $G=SU(2)$ in codimension one.
In Weierstrass form this is achieved by setting
\bea \label{IImodelWeier1}
f = z_1 \, f_0, \qquad g= z_1^{\mu_g} \, g_0 \quad {\rm for} \quad \mu_g \geq2  \qquad \rightarrow \quad  \Delta = z_1^3 \, (4 f_0^3 + 27  z_1^{2 \mu_g - 3} \, g_0^2).   
\eea
The intersection points $P_1 = \Sigma_1 \cap \Sigma_0$, with $\Sigma_0: 4 f_0^3 + 27  z_1^{2 \mu_g - 3} \, g_0^2=0$, lie at $z_1 = f_0 = 0$.
This gives rise to the following vanishing orders at $P_1$:
\be \label{TypeIIImodelordersP1}
\ba
&\mu_g= 2: \qquad &{\rm ord}(f,g,\Delta)|_{P_1} = (2,2,4), \qquad  {\rm III} \rightarrow {\rm   IV}, \cr
&\mu_g\geq 3: \qquad &{\rm ord}(f,g,\Delta)|_{P_1} = (2,\mu_g,6),  \qquad  {\rm III} \rightarrow  {\rm I}^*_0,
\ea
\ee
where we also indicate the naively expected fiber type at the enhancement points from Kodaira's table. 

As will be shown in detail in Section \ref{IIIresolution}, for $\mu_g =2$ and $\mu_g =3$ a crepant resolution of the Weierstrass model exists which in particular completely resolves the codimension-two fibers over $P_1$. For $\mu_g =2$ the resolved fiber is indeed of type IV, while for $\mu_g =3$ it has three $\mathbb P^1$s deleted compared to the naively expected standard Kodaira fiber ${\rm I}^*_0$.
After performing this resolution one finds \emph{two}  localised hypermultiplets per enhancement point $P_1$ in representation ${\bf 2}$ of the gauge algebra $SU(2)$ from wrapped M2-branes wrapping suitable fibral curves. The appearance of two such hypermultiplets per point (as opposed to just one) is quite interesting by itself and discussed at the end of section \ref{IIIresolution}.

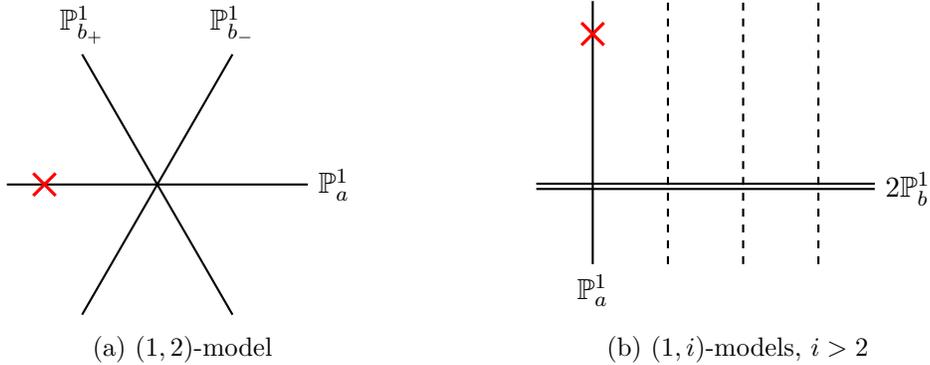
\begin{figure}
\centering
\begin{subfigure}[b]{0.45\textwidth}
	\centering
    

\begin{tikzpicture}[thick]
\draw (-2,0) -- (2,0) node [right] {$\mathbb P^1_{a}$} node [very near start,draw,cross out,very thick, red] {};
\draw[cm={cos(60) ,-sin(60) ,sin(60) ,cos(60), (0,0)}] (-2,0) node [above] {$\mathbb P^1_{b_+}$} -- (2,0);
\draw[cm={cos(60) ,sin(60) ,-sin(60) ,cos(60), (0,0)}] (-2,0) -- (2,0) node [above] {$\mathbb P^1_{b_-}$};
\end{tikzpicture}

    \caption{$(1,2)$-model}
\end{subfigure}
~
\begin{subfigure}[b]{0.45\textwidth}
	\centering

    \begin{tikzpicture}[thick]
\coordinate (a) at (1,0);
\coordinate (b) at (2,0);
\coordinate (c) at (3,0);
\coordinate (d) at (4,0);
\coordinate (len) at (0,3.5);
\coordinate (end) at (4.75,1);
\coordinate (shift) at (0,.07);

\draw (a) node [below] {$\mathbb P^1_a$} -- ($(a) + (len)$) node [very near end,draw,cross out,very thick, red] {};
\draw [dashed] (b) -- ($(b) + (len)$);
\draw [dashed] (c) -- ($(c) + (len)$);
\draw [dashed] (d) -- ($(d) + (len)$);
\draw (0.25,1) -- (end) node [right] {$2 \mathbb P^1_b$};
\draw ($(0.25,1) + (shift)$) -- ($(end) + (shift)$);

\end{tikzpicture}

    \caption{$(1,i)$-models, $i>2$}
\end{subfigure}
\caption{Affine Dynkin diagram of the partially resolved fiber over $P_1: z_1 = f_0 =0$. The red cross denotes the intersection with the zero-section $z=0$ of the Weierstrass model and the dashed lines symbolize the deleted $\mathbb P^1$s in the standard Kodaira fiber which are not realized in the actual fiber. The resolution was obtained as resolution of a Tate model realising the vanishing orders (\ref{IImodelWeier1}).}
\label{fig:12z1a4-1}
\end{figure}

By contrast, for certain values $\mu_g \geq 4$, the singularity at $P_1$ turns out to be $\mathbb Q$-factorial terminal. Concretely we have studied $\mu_g = 4$, 
$\mu_g = 5$, $\mu_g = 7$. In these cases the (partial) resolution $\hat Y_3$ presented in Section \ref{IIIresolution}  yields a monodromy reduced  I$_0^*$ fiber, depicted in figure \ref{fig:12z1a4-1}, with a residual terminal hypersurface singularity of type
\bea \label{RHMtypeIII1}
z^a + x_1^2 + x_2^2 + x_3^2 = 0
\eea
with
\bea
&a = 2 \qquad {\rm for} \, &\mu_g = 4, \label{localhyperIII4}\\
&a = 3 \qquad {\rm for} \, &\mu_g = 5 ,\label{localhyperIII5}\\
&a=5   \qquad {\rm for} \, &\mu_g = 7 \label{localhyperIII7}. 
\eea

Note that the conifold singularity for $\mu_g = 4$ is indeed $\mathbb Q$-factorial terminal due to higher order terms; hence a small resolution does not exist.
The reason for the appearance of residual terminal singularities is the localisation of  {\it uncharged} hypermultiplets at these points, in addition to the two localised hypermultiplets in representation {\bf 2} of $SU(2)$ present also in the resolvable cases with $\mu_g =2,3$. Since the residual terminal singularity is either a $\mathbb Q$-factorial terminal double point (for $\mu_g =4$) or of  odd Kleinian type (for $\mu_g = 5 ,7$), our general results of Section \ref{sec_defs} can be applied. In particular, the number of localised uncharged hypermultiplets per point $P_1$ is given by $m_{P_1} =1$ for $\mu_g = 4$, $m_{P_1} = 2$ for $\mu_g = 5$ and by $m_{P_1} = 4$ for $\mu_g = 7$. 
Furthermore, we can follow the programme of computing the total number of uncharged hypermultiplets via (\ref{loc-nonlocHyerps}). 
The crucial step is again to evaluate $\chi_{\rm top}(\hat Y_3)$, where $\hat Y_3$ denotes the partial resolution. 
As can be deduced from the explicit form of the fiber in Figure \ref{fig:12z1a4-1} following the procedure in appendix \ref{sec:howToComputeChiP},  the partially resolved I$_0^*$ fiber over $P_1$ contributes 
\bea\label{chitopI0*}
\chi_{\rm top}(X_{P_1}) = (2-1) + (2-1) + 1 = 3.
\eea
Furthermore, the value of the parameter $\epsilon_1$ correcting for the singularities of $\Sigma_0$ at $P_1$ follows readily from (\ref{eq:epsilonFormulae}) as 
\bea
\epsilon_1 = 4 \mu_g -9.
\eea
 Indeed $\Sigma_0$ at $P_1$ takes the local form
$x^3 +y^{2\mu_g -3}=0$. If we specialise the base to $B_2 = \mathbb P^2$ and let $\Sigma_1$ wrap the curve $z_1=0$ as before, we find the values summarized\footnote{The corresponding values for the two smooth models with $(\mu_f, \mu_g) = (1,2)$ and $(1,3)$ are listed in table \ref{tab:oneBraneResults-smooth} in appendix \ref{app:toricResolutionsOneBrane}. } in table \ref{tab:oneBraneResultsnonResGauge} for $\chi_{\rm top}(\hat Y_3)$ and correspondingly for the numbers of localised and unlocalized uncharged hypers. 
The spectrum is indeed consistent with the cancellation of all gravitational and gauge anomalies.

\begin{table}
\centering
\begin{tabular}{lrrr}
\toprule
$(\mu_f ,\mu_g)$    & $(1,4)$ & \hspace{3em}$(1,5)$ &\hspace{3em} $(1,7)$\\
$m$                 & $3$     & $3$     & $3$     \\\midrule
Enhancements        & III $\to \text I_0^*$& III $\to \text I_0^*$ & III $\to \text I_0^*$\\
Gauge Group         & $SU(2)$ & $SU(2)$ & $SU(2)$\\
$n_V=\dim (G)$      & $3$     & $3$     & $3$    \\
$\mathrm{rk}(G)$    & $1$     & $1$     & $1$    \\
$h^{1,1} (\hat Y_3)$ & $3$  & $3$     & $3$    \\
$B_1$               & $11$    & $11$    & $11$   \\
$\ki(X_{P_1})$      & $3$     & $3$     & $3$    \\
$\epsilon_1$        & $7$     & $11$    & $19$   \\\midrule
$a$                 & 2       & 3       & 5      \\
$m_P$               & 1       & 2       & 4      \\
$\ki (\hat Y_3)$         & $-445$  & $-434$  & $-412$  \\\midrule
non-loc neutral hypers $n^0_{H,n-l}$ & $221$& $210$ & $188$ \\
localised neutral  hypers $n^0_{H,l}$ & $11$& $22$ & $44$    \\\midrule
charged hypers $n^c_{H}$& $44$& $44$    & $44$    \\
Representation      & $2\times$fund.& $2\times$fund.& $2\times$fund.\\\midrule
$273-(n^0_{H,n-l}+n^0_{H,l}+n^c_{H}-n_V)$ & 0   & 0    & 0    \\\bottomrule
\end{tabular}
\caption{Non-resolvable models with non-trivial gauge group. The parameter $a$ characterises the form of the codimension-two singularity, given by $z^a+ x_1^2+x_2^2+x_3^2=0$, and $m_P$ denotes the corresponding Milnor number.}
\label{tab:oneBraneResultsnonResGauge}
\end{table}

\subsection{Partial Resolution} \label{IIIresolution}

The specific form of the fiber at $P_1$ and the local expression (\ref{RHMtypeIII1}) for its terminal singularities can be achieved by a patchwise partial resolution of the most generic Weierstrass model subject to the restriction (\ref{IImodelWeier1}).
As an alternative we realize (\ref{IImodelWeier1}) as a Tate model as this allows us to construct a global resolution by a blowup of the toric ambient space into which hypersurface is embedded. 
In the remainder of this section we present this model and its (partial) resolution. 
In order to reproduce (\ref{IImodelWeier1}) the sections $a_i$ in the Tate model (\ref{eq:TateForm}) must be restricted as
\bea
a_i =  \tilde a_i \, z_1^{k_i},
\eea
for the values of $k_i$  as  collected in table~\ref{tab:vanishingOrders}.  That this indeed leads to the desired vanishing orders for $f$ and $g$ can be checked via (\ref{fgfrom Tate}).
\begin{table}\centering
\begin{tabular}{lrrrrr}
\toprule
$(\mu_f,\mu_g)$ & $k_1$& $k_2$& $k_3$& $k_4$& $k_6$\\\midrule
$(1,2)$ & 1&1&1&1&2\\
$(1,3)$ & 1&2&2&1&3\\
$(1,4)$ & 2&3&2&1&4\\
$(1,5)$ & 2&4&3&1&5\\
$(1,7)$ & 3&6&4&1&7\\
\bottomrule
\end{tabular}
\caption{Vanishing orders $k_i$ of the Tate sections $a_i$ along the locus $\Sigma_1$. 
}
\label{tab:vanishingOrders}
\end{table}
To resolve the  singularity in the fiber over $\Sigma_1$, located at $x=y=z_1=0$, we perform the blow-up 
\bea
x \to e_1x, \quad y \to e_1y, \quad z_1 \to e_0 \, e_1.
\eea
The proper transform $PT$ of the Tate equation is then a 
hypersurface in a toric fiber ambient space with coordinates $x,y,z,e_0,e_1$ and toric weights displayed in table \ref{tab:ScalingRelations}. However, the hypersurface as such is not the most general hypersurface compatible with these scaling relations. 
This most generic hypersurface would rather give rise to  Kodaira fibers of type $\mathrm I_2$ in codimension one, and not type III. In particular, the dual polytope does not reproduce the monomials in $PT$. In this sense, this type III model cannot be analysed via the technology of tops \cite{Candelas:1997eh,Bouchard:2003bu}.

Nonwithstanding this fact  we can still compute the Stanley-Reisner ideal (SRI) of the toric ambient space and analyse the hypersurface $PT$ by hand. 
There exist two triangulations of the ambient space, and for concreteness we choose the triangulation with SRI given by
\begin{align}
{\rm SRI}  =  \langle ze_1,xyz,xye_0 \rangle \, .
\end{align}
Most of the potentially singular loci of our models where $PT = dPT=0$ are excluded by the SRI. However, there remain some singularities which are displayed in table~\ref{tab:singLoci}. Note that in the $(1,3)$-model the codimension of the singular locus is too high and therefore non-existent on Calabi-Yau threefolds, which is the case of interest in this paper.
For $\mu_g \geq 4$, by contrast, the blow-up defines only a {\it partial} Calabi-Yau resolution $\hat Y_3$. 
In fact, the remaining singularity is immediately identified as a terminal hypersurface singularity of the type advocated in (\ref{localhyperIII4}), (\ref{localhyperIII5}), (\ref{localhyperIII7}): Near the singularity, we can set $z_1=e_0=\tilde a_i=1$ for $i \neq 4$ and write the proper transform as 
\bea
PT = e_1^{k_6-2} + x ( \tilde a_4 + e_1^{k_2} \, x  + e_1 x^2 - e_1^{k_1} y) - y (e_1^{k_3-1} + y),
\eea
with $k_i$ the vanishing orders displayed in table \ref{tab:vanishingOrders}. The claim then follows by completing the square in $y$,  keeping only the leading monomials in $e_1$, and  by a simple coordinate change.

To analyse the structure of the fibers, and in particular to compute $\chi_{\rm top}$ for the critical fibers over $P_1$, note first that 
the fiber over $z_1=0$ consists of two rational curves, given by the vanishing of $e_0$ and $e_1$, respectively. Concretely,
\begin{align*}
\p_A:\: PT|_{e_0 \to 0} &= y^2-e_1 x^3,\\
\p_B:\: PT|_{e_1 \to 0} &= \begin{cases}
y^2- \tilde a_4 e_0\, x z^4 +\tilde a_3 e_0 \,yz^3  -\tilde a_6 e_0^2\, z^6 & \text{for} \, \,  (\mu_f,\mu_g) = (1,2)\\
y^2-\tilde a_4e_0 \, xz^4 & \text{for} \, \,  (\mu_f,\mu_g) = (1,i), \,  i = 3,\ldots,7.
\end{cases}
\end{align*}
These two equations do not factorize and therefore define two $\p$s in the fiber called $\p_A$ and $\p_B$. They intersect at $\ib{e_0} \cap \ib{e_1} \cap \ib{y} $ with order two and thus realise a type III fiber (see figure \ref{fig:fiberTypes}). Note that $\p_A$ is intersected by the zero-section of the Weierstrass model.
The charge of $\p_B$ under the Cartan of $SU(2)$ is given by minus the intersection product of the curve $\p_B$ and the divisor $E_1$ locally defined by $e_1=0$,
\begin{align}
- \p_B \circ E_1 &= - [e_1] \cdot \left[y^2+\ldots\right] \cdot [e_1] = [e_1] \cdot [ y^2 ] \cdot [e_0] = 2,
\end{align}
where we used that $[e_1] = -[e_0]$ (see table~\ref{tab:ScalingRelations}).
This identifies $\p_B$ as the simple root of $SU(2)$ as expected.

\begin{table}
 \centering
\begin{tabular}{lccccccc}
\toprule
& $x$ & $y$ & $z$ & $e_0$ & $e_1$ \\\midrule
$[Z]$ & 2&3&1&$\cdot$&$\cdot$\\
$[E_1]$&$-1$&$-1$&$\cdot$&$-1$&$1$\\\bottomrule
\end{tabular}
\caption{Fiber type III ($SU(2)$)} \label{tab:ScalingRelations}
\end{table}

From the discriminant of the Tate models,
\begin{align}
\Delta &= 
\tfrac1{16} \,z_1^3 \cdot \left( 64\, \tilde a_4^3 + \mathcal O (z_1) \right),
\end{align}
we read off that the fiber type enhances at $z_1 =\tilde a_4 =0$, which corresponds to the location of the points $P_1$.
 In the $(1,2)$-model the fiber over this point takes the following form,
\begin{align*}
PT|_{e_0,\tilde a_4\to0} &= \underbrace{y^2-e_1x^3}_{\p_a},\\
PT|_{e_1,\tilde a_4\to0} &= \underbrace{y^2+\tilde a_3 e_0 \,yz^3 -\tilde a_6 e_0^2\, z^6}_{\p_B \text{ at } \tilde a_4 \to 0}\\
&= \underbrace{\left(y+\tfrac12 \tilde a_3e_0\,z^3 + \sqrt{\tilde a_3^2+4\tilde a_6}\,e_0\,z^3 \right)}_{\p_{b_+}} \cdot \underbrace{\left(y+\tfrac12\, \tilde a_3e_0\,z^3 - \sqrt{\tilde a_3^2+4\tilde a_6}\,e_0\,z^3 \right)}_{\p_{b_-}}.
\end{align*}
At the codimension-two enhancement locus the curve $\p_B$ therefore splits into two curves, called $\p_{b_+}$ and $\p_{b_-}$. All three curves meet at $\ib{e_0} \cap \ib{e_1} \cap \ib{y}$ with multiplicity one, corresponding to a Kodaira fiber of Type IV as shown in figure~\ref{fig:fiberTypes}.  This is in agreement with the vanishing orders of $f$, $g$ and $\Delta$. For the $(1,i)$-models with $i>2$ the situation is slightly different:
\begin{align*}
\p_a:\: PT|_{e_0,\tilde a_4\to0} &= y^2-e_1x^3,\\
2\, \p_b:\: PT|_{e_1,\tilde a_4\to0} &= y^2.
\end{align*}
The second equation now describes a non-reduced curve of multiplicity two. 
We denote the reduced curve defined by $e_1=\tilde a_4 = y=0$ as $\p_b$.
Taking this into account, the fiber can indeed be interpreted as an I$_0^*$ Kodaira fiber with three multiplicity-one curves deleted, again in agreement with the vanishing orders (\ref{TypeIIImodelordersP1}). The contribution of such a fiber to the topological Euler characteristic has already been computed in (\ref{chitopI0*}).

 \begin{table}
\centering
\begin{tabular}{llr}\toprule
$(\mu_f, \mu_g)$ & Singular locus after resolution & $a$\\\midrule
$(1,2)$ & $\varnothing$ & --\\
$(1,3)$ & $\ib{e_1} \cap \ib{\tilde a_6} \cap \ib{\tilde a_4} \cap \ib x \cap \ib y = \varnothing$ & --\\
$(1,4)$ & $\ib{e_1} \cap \ib{\tilde a_4} \cap \ib x \cap \ib y$ & 2\\
$(1,5)$ & $\ib{e_1} \cap \ib{\tilde a_4} \cap \ib x \cap \ib y$ & 3\\
$(1,7)$ & $\ib{e_1} \cap \ib{\tilde a_4} \cap \ib x \cap \ib y$ & 5\\
\bottomrule
\end{tabular}
\caption{Singular locus of proper transforms describing the Calabi-Yau threefold $\hat Y_3$. The singularity parameter $a$ describes the local form of the singularity: $z^a + x_1^2+x_2^2+x_3^2$.}
\label{tab:singLoci}
\end{table}

Next, we need the Cartan weights of $\p_{b_\pm}$ and $\p_b$. The curve $\p_a$ is still intersected by the zero-section and does not play a role in our analysis. We compute
\begin{align*}
 - \p_b \circ E_1 &=  - [e_1] \cdot [y] \cdot [e_1]  =  \cdot [e_0] \cdot [y] \cdot [e_1] =1,\\
 - \p_{b_\pm} \circ E_1 &= -  [e_1] \cdot \left[ y+\tfrac12 \tilde a_3e_0\,z^3 \pm \sqrt{\tilde a_3^2+4\tilde a_6}\,e_0\,z^3  \right] \cdot [e_1] = [e_1]\cdot [y] \cdot [e_0] = 1.
\end{align*}
The situation is hence completely equivalent in both cases. The $SU(2)$ root $\alpha$ is given by $\p_B$ which has factorized into $\p_{b_+} + \p_{b_-}$ in the $(1,2)$-case. The highest weight of the fundamental representation of $SU(2)$ is $w=1$ which is either represented by $\p_{b_+}$ or $\p_{b_-}$ (or by $\p_b$). In this fashion we can build:
\begin{align*}
w = \p_{b_+}:\quad &\begin{pmatrix}w-\alpha\\w\end{pmatrix} = \begin{pmatrix}\p_{b_+}-(\p_{b_+}+\p_{b_-})\\\p_{b_+} \end{pmatrix} = \begin{pmatrix}-\p_{b_-}\\\p_{b_+} \end{pmatrix} = \begin{pmatrix}-1\\1\end{pmatrix},\\
w = \p_{b_-}:\quad &\begin{pmatrix}w-\alpha\\w\end{pmatrix} = \begin{pmatrix}\p_{b_-}-(\p_{b_+}+\p_{b_-})\\\p_{b_-} \end{pmatrix} = \begin{pmatrix}-\p_{b_+}\\\p_{b_-} \end{pmatrix} = \begin{pmatrix}-1\\1\end{pmatrix}.
\end{align*}
Clearly both pairs of fibral curves each describe the weight vector of a  fundamental representation ${\bf 2}$ of $SU(2)$, but they are not independent as they only differ by a minus sign.
An M2-brane can wrap each fibral curve with two orientations, and the sign difference above can be translated into the orientation of the wrapped M2.
This way one would naively conclude that there is one massless hypermultiplet in the ${\bf 2}$ of $SU(2)$ localised at the enhancement point. However, as observed already in \cite{Grassi:2011hq}, 
near the enhancement point the Weierstrass equation (with $z \equiv 1$) can be written as 
\be \label{weierIII}
\ba
P_W & =  - y^2 + x^3 + f_0 z_1 x + z_1^2 + {\cal O}(z_1^3)  \\
&= -y^2 + x^3  - \frac{1}{4} f^2_0 x^2 + (z_1 + \frac{1}{2} f_0 x)^2 +  {\cal O}(z_1^3).
\ea
\ee
The enhancement point at  $f_0 = z_1 =0$ corresponds to an $A_2$ singularity $-y^2 + x^3 + z_1^2 = 0$. The Weierstrass equation (\ref{weierIII}) is a deformation of this $A_2$ to an $A_1$ singularity with deformation parameter $t = \frac{1}{4} f^2_0$.
Since this parameter appears quadratically, the number of massless hypermultiplets in representation {\bf 2} of $SU(2)$ per enhancement point is given by $2$ \cite{Katz:1996xe,Grassi:2011hq} (see table~\ref{tab:oneBraneResultsnonResGauge}).\footnote{Another way to see this is by interpreting the Type III model as a specialization of an ${\rm I}_2$ model in which two enhancement points ${\rm I}_2 \rightarrow {\rm I}_3$, each carrying one hypermultiplet in the ${\bf 2}$ of $SU(2)$, coalesce to one ${\rm Type \, III}  \rightarrow {\rm Type \, IV }$ point.}
This number is also required by cancellation of all gauge and gravitational anomalies.

\section{Conclusions and Outlook} \label{sec:Conclusions}

We have investigated non-crepant resolvable codimension-two singularities in F-theory compactifications. 
The physical interpretation of such singularities is very simple: Massless matter uncharged under any gauge group localizes in the fiber and cannot be rendered massive along a supersymmetric direction in the Coulomb branch, as would be required for a Calabi-Yau resolution to exist.
For a certain class of isolated terminal $\mathbb Q$-factorial hypersurface singularities (of Kleinian type $A_{a-1}$ with $a=2$ or $a$ odd) on elliptic Calabi-Yau threefolds we have shown how to compute both the unlocalised and the localised uncharged hypermultiplets  in terms of the Milnor number of the singularity and the topological Euler characteristic of the singular variety.  
These expressions have been put to a successful test in a number of examples and indeed produced an anomaly-free spectrum in the associated six-dimensional F-theory compactification. 
The methods hold for more general singularities,  which do not appear as examples in this paper \cite{ArrasGrassiWeigandM}.

 There are many interesting directions for future investigations. An obvious question concerns codimension-two terminal singularities on elliptic Calabi-Yau fourfolds:
 The physical intuition about the appearance of massless matter is independent of the dimension of the compactification. Thus we clearly expect similarly localised uncharged hypermultiplets, now from terminal singularities in the fiber over a curve $C$ in base. 
 Extrapolating the counting in \cite{Intriligator:2012ue,Bies:2014sra} of massless matter in F-theory on smooth fourfolds 
 a natural conjecture for the number of localised vectorlike pairs of $N=1$ multiplets would be     
\bea
n =  g(C) \, m_P 
\eea
with $m_P$ the Milnor number of the singularity in the fiber and $g(C)$ the genus of $C$.
This formula would hold in the absence of $G_4$-flux. Given the nature of the uncharged states as part of the complex structure moduli of the singular variety we expect the counting to be modified at best by horizontal fluxes $G_4 \in H^{2,2}_{\rm hor}(Y_4)$, as opposed to fluxes in the vertical part of the middle cohomology. 
Clearly the investigation of the middle cohomology of singular fourfolds might become equally challenging and exciting, both from a conceptual and a computational viewpoint. 

While the terminal singularities studied in this paper do not allow for a crepant resolution, the singular varieties can be resolved into a non-Calabi-Yau space.
In the spirit of Section \ref{subsec_non-crepantb},  in the dual M-theory the resolution breaks supersymmetry by moving along an obstructed direction in the Coulomb branch.
Nonetheless certain physical quantities might be robust enough to be computed from this non-Calabi-Yau phase. Ideas along these lines have been put forward in \cite{Grimm:2010ez,Grimm:2011tb,Braun:2014nva}.
Some of the examples studied in this paper will involve non-flat non-Calabi-Yau resolutions, and a preliminary analysis suggests that the counting of localised uncharged states is indeed reproduced. We look forward to reporting on these questions in future work.

\vspace{5pt}

\noindent{\bf  Acknowledgements}

We thank Paolo Aluffi, Markus Banagl, Jim Halverson, Craig Lawrie, Ling Lin, Laurentiu Maxim, Christoph Mayrhofer, Dave Morrison, Eran Palti, Michele Rossi, Sakura Sch\"afer-Nameki, Julius Shaneson, Vasudevan Srinivas and Washington Taylor for discussions. AG and TW thank the Aspen Center for Physics and the Fields Institute, Toronto,
for hospitality during part of this project.
The work of TW is partially supported by DFG under TR33 'The Dark Universe'.

\appendix

\section{Computation of $\ki(\hat Y_3)$} \label{app-chi}

In this appendix, we summarize, for the reader's convenience, the computation of the 
topological  Euler characteristic of an elliptically fibered Calabi-Yau threefold  as presented in \cite{Grassi:2000we}, and explain its use in the presence of terminal singularities over codimension-two points in the base.
Our notation for the (in general singular) Weierstrass model of an elliptically fibered Calabi-Yau threefold $Y_3$ has already been introduced in Section \ref{sec_Ftheorynotation}.  For simplicity, we first assume, as in \cite{Grassi:2000we}, that the gauge group does not contain any abelian factors and that it has only one semi-simple factor, see figure \ref{fig:chi}. Later, in Section~\ref{cha:twoBranes}, we will generalise the computation for $\ki$ to theories with two identical simple factors.

An important point to note is that the topological Euler characteristic we compute is then the Euler characteristic of a (partial) Calabi-Yau resolution $\hat Y_3$ of this Weierstrass model.
If all singularities are crepant resolvable, as was the case for the geometries studied in  \cite{Grassi:2000we}, $\hat Y_3$ is a smooth Calabi-Yau space. 
In the presence of terminal singularities it is understood that $\hat Y_3$ is a smooth Calabi-Yau with at worst those terminal singularities in codimension-two which cannot be resolved completely in a crepant way.

The topological Euler characteristic $\ki(\hat Y_3)$ has two important properties. First, one can split the space into smaller ones and compute their Euler characteristic and then sum all contributions up. This is possible due to the Mayer-Vietoris sequence. Second, for a product space the topological Euler characteristic can be expressed as a product of the Euler characteristics of the factors. In our case the elliptic fibration is locally a product space and therefore we can compute the Euler characteristic of the base and multiply it with the Euler characteristic of the fiber. But since the fibration is generally non-trivial this is only possible locally.

We can split the total space into five components:
\begin{enumerate}
	\item The fibers   $ \bigcup_i \pi^{-1} (P_i)$  over the intersection points $P_i$ contribute
	\begin{align}\label{eq:firstContribution}
	\sum_i \ki (X_{P_i}) \cdot B_i \, ,
	\end{align}
	where $B_i$ denotes the number of points $P_i$.
	\item The generic fiber over $\Sigma_1$: $\pi^{-1} (\Sigma_1 \setminus \bigcup_i P_i)$ contributes as follows: The Euler characteristic of $\Sigma_1$ without the enhancement points is given by $\ki (\Sigma_1) = 2-2g(\Sigma_1)$ minus the number of points $P_i$, $\sum_i B_i$. It must be multiplied by the Euler characteristic of the fiber, and contributes
	\begin{align}\label{eq:secondContribution}
	\ki(X_{\Sigma_1}) \cdot \Big( 2-2\,g(\Sigma_1) - \sum_i B_i \Big).
	\end{align}
	\item Analogously, there is a contribution from the general fiber over $\Sigma_0$ given by
	\begin{align}\label{eq:thirdContribution}
	\ki \Big( \pi^{-1} (\Sigma_0 \setminus Q \setminus \bigcup_i P_i) \Big).
	\end{align}
	\item Finally, the fibers over the cuspidal points $Q$ contribute
	\begin{align}\label{eq:fourthContribution}
	\ki(X_Q) \cdot C,
	\end{align}
	where $C$ is the number of such points. 
	
	\item The general fiber over points where the discriminant does not vanish does not contribute to the Euler characteristic since the $\ki$ of a torus is zero.
\end{enumerate}

Let us compute the  different contributions in turn. \eqref{eq:firstContribution} is already in its final form.
The fiber $X_{P_i}$ is in general given by a chain of rational curves intersecting each other at points. 
We stress again that if the Weierstrass model $Y_3$ has non-crepant resolvable singularities, then  $X_{P_i}$  is the fiber in a partial Calabi-Yau resolution $\hat Y_3$ of $Y_3$ in which all codimension-one singularities are resolved and only the terminal singularities in codimension-two remain. 
The computation of the topological Euler characteristic of a chain of $\mathbb P^1$s is standard, and  we review it in Section \ref{sec:howToComputeChiP}. 
In particular, the results of this computation show that $\ki (X_{\Sigma_1}) = m$ as $X_{\Sigma_1}$ is of standard Kodaira fiber type.
Similarly, in \eqref{eq:fourthContribution} we must set $\ki(X_Q) = 2$ for the cuspidal fibers (type II in Kodaira's  classification). What remains is to compute the number of cusps of $\Sigma_0$ and to bring the third contribution into a nicer form.

\paragraph{The number of cusps $C$.}
Cusps appear as soon as both $f$ and $g$ vanish along $\Sigma_0 \setminus \bigcup_i P_i$. Since $f$ and $g$ are of the form \eqref{eq:fandg} only $f_0$ and $g_0$ can vanish along $\Sigma_0$ away from $\Sigma_1$. The number of intersection points is naively $(-4 K_B - \mu_f \Sigma_1)\cdot(-6K_B - \mu_g \Sigma_1)$. However, we have to correct it by the intersection multiplicity $\mu_{P_i} (f,g)$ of $f_0$ and $g_0$ at the points $P_i$. All in all, the number of cusps is given by
\begin{align}
C = 24 K_B^2 + \Big(4\mu_g  + 6 \mu_f  \Big) K_B \cdot \Sigma_1 + \mu_f  \mu_g  \Sigma_1^2 - \sum_i \mu_{P_i}(f,g) B_i .
\end{align}

\paragraph{The third contribution.}
Finally, to compute $\ki \left(\pi^{-1} (\Sigma_0 \setminus Q \setminus \bigcup_i P_i ) \right)$, note first  that along $\Sigma_0$ (away from $Q$ and $P_i$) the vanishing order structure is $(f,g,\Delta)|_{\Sigma_0} = (0,0,1)$, i.e. the fiber develops a type ${\rm I}_1$ singularity with Euler characteristic $\ki (X_{\Sigma_0}) = 1$.

If $\Sigma_0$ was a smooth curve, its topological Euler characteristic would be given by $-(K_B + \Sigma_0) \cdot \Sigma_0$. However,  $\Sigma_0$ may not be smooth at the intersection points $P$, and it is definitely not smooth at the (cuspidal) self-intersection points $Q$. Besides, one has to exclude all these points because they are already taken into account in \eqref{eq:firstContribution} and \eqref{eq:fourthContribution}. Therefore, we have to include correction terms for each point $P_i$ and $Q$,
\begin{align}
\ki \Big(\pi^{-1} ( \Sigma_0 \setminus Q \setminus \bigcup_i P_i  ) \Big) = \Big( -(K_B + \Sigma_0) \cdot \Sigma_0 + \sum_i \epsilon_i B_i + \epsilon_c C \Big) \cdot \underbrace{\ki (X_{\Sigma_0})}_{=1}.
\end{align}
Obviously, $\epsilon$ has to be defined such that it is $-1$ if the considered point is a smooth point on the curve.

To evaluate $(K_B + \Sigma_0) \cdot \Sigma_0 $ note that the 
 Calabi-Yau condition for an elliptic fibration $\Delta \in \mathcal O (-12 K_B)$ implies $\Sigma_0 = -12 K_B - m \Sigma_1$. It immediately follows that 
 \bea
 \Sigma_0 \cdot \Sigma_1 = -12 K_B \cdot \Sigma_1 - m \Sigma_1^2.
 \eea
 Applying these two relations several times gives
\begin{align}
-(K_B + \Sigma_0) \cdot \Sigma_0 = -132\, K_B^2 + m \,K_B \cdot \Sigma_1 + 2 m \,\Sigma_0 \cdot \Sigma_1 + m^2 \,\Sigma_1^2 .
\end{align}

Finally, let us properly define the correction factors $\epsilon_i$. As noted already, for a smooth point $\epsilon$ must take the value $-1$. Let us consider a curve $D$ with singular point $P\in D$. $\phi_1: B_1 \to B$ shall be the blow-up of the point $P$ with exceptional divisor $E$. We define the quantity $\alpha_1$ via $D_1 = \phi_1^* (D) - \alpha_1 (P) E$ where $D_1$ is the strict transform of the curve.\footnote{
	To illustrate the definition of $\alpha_1$ let us look at a simple example: Let the curve $D$ be given by the equation $x^3+y^3=0$. It is singular at $(0,0)$. The blow-up $x\to xy, y\to y$ leads to $y^3\cdot(x^3+1)=0$. Then $y$ is the exceptional divisor of the blow-up and appears with multiplicity $\alpha_1 = 3$. 
}
Then the Euler characteristic of the blown-up curve is
\begin{align} \label{kD1}
\ki (D_1) = -( K_{B_1} + D_1 ) \cdot D_1 = -  (K_B + D)\cdot D - \alpha_1 (P) \cdot (\alpha_1 (P) - 1).
\end{align}

We perform successive blow-ups until the point $P$ is smooth. Since we do not want to include the singular point itself in our calculation of $\ki$ we have to subtract the number of preimages of $P$ under the total blow-up $\phi$. We combine the total correction of $\ki$ due to the singular point $P$ into the definition of $\epsilon$,
\begin{align} \label{epsilonP1}
\epsilon_P := \sum_{i} \alpha_i (P) \cdot  \Big(\alpha_i (P)  - 1\Big) - \# \phi^{-1} (P),
\end{align}
where $i$ runs over the successive blow-ups one has to perform until the singularity is smoothed out completely.\footnote{
	In our above example $\# \phi^{-1} (P)  = 3$ and therefore $\epsilon = 3\cdot 2 - 3 = 3$. Let us look at another example: $x^3+y^5=0$. The first blow-up is $x\to xy$: $y^3 \cdot (x^3+y^2) = 0$. So $\alpha_1 = 3$. Then perform $y\to xy$: $x^2 \cdot (x+y^2) = 0$. Since $x+y^2$ has only one solution $\# \phi^{-1} (P) = 1$ and $\epsilon = 3 \cdot 2 + 2 \cdot 1 - 1 = 7$. 
}
In Section~\ref{sec:howToComputeEpsilon} we will explicitly compute $\epsilon_i$ for all singularity types which appear in this paper. 

\paragraph{Final Result.} Putting everything together we arrive at the final expression \cite{Grassi:2000we} of the topological Euler characteristic of an elliptically fibered Calabi-Yau threefold $\hat Y_3$ over $B_2$ with singular locus $\Sigma_0 \cup \Sigma_1$ as specified above 
\begin{align}\label{eq:chi}
\begin{split}
\ki (\hat Y_3) &=  \ki \Big(\bigcup_i \pi^{-1} (P_i) \Big)  + \ki \Big(\pi^{-1} (\Sigma_1 \setminus \bigcup_i \pi^{-1} (P_i) ) \Big) +  
  \ki \Big(\pi^{-1} (\Sigma_0 \setminus Q \setminus \bigcup_i \pi^{-1} (P_i)) \Big)  \\
  & \quad + \ki \Big(\pi^{-1} (Q) \Big)\\
&= \Big( \sum_i B_i \cdot \ki (X_{P_i}) \Big) + m\, \Big(2-2g-\sum_i B_i\Big) \\
&\quad - 132 K_B^2 + m \,K_B \cdot \Sigma_1 + 2 m \,\Sigma_0 \cdot \Sigma_1 + m^2 \,\Sigma_1^2  + 3C+ \sum_i \epsilon_i B_i ,
\end{split}
\end{align}
with 
\begin{align}\label{eq:chi2}
C = 24 K_B^2 + \Big(4\mu_g  + 6 \mu_f  \Big) K_B \cdot \Sigma_1 + \mu_f  \mu_g  \Sigma_1^2 - \sum_i\mu_{P_i}(f,g) B_i.
\end{align}

\subsection{Computation of  $\epsilon_P$}\label{sec:howToComputeEpsilon}

In this article  we encounter three different classes of points for which we have to evaluate the parameters $\epsilon_P$ defined in (\ref{epsilonP1}): smooth points, singularities of the form $x^2+y^n=0$ for $n\geq 2$ and singularities of the form $x^3+y^n=0$ for $n\geq 3$. We now derive the general expression
\begin{align}\label{eq:epsilonFormulae}
\epsilon_{\mathrm{smooth}} = -1, \quad\quad \epsilon_{x^2+y^n,\: n\,\geq\, 2} = n-2, \quad\quad \epsilon_{x^3+y^n,\: n\,\geq\, 3} = 2n-3.
\end{align}
for $\epsilon_P$ in each case. Since all singularities are located at $(x,y) = (0,0) $ we have suppressed the index $P$. Additionally, we drop the index $i$ which counts the number of successive blow-ups because it should be clear from the context.

\paragraph{Smooth points:}
	As we already pointed out there is no need to blow up smooth points ($\alpha = 0$ and $\# \phi^{-1} (P) = 1 $). Thus, $\epsilon = \alpha (\alpha - 1) - \# \phi^{-1} (P) = -1$. 

\paragraph{${\bf x^2+y^n=0}$ for $n\geq 2$: }
We prove by induction: First consider the curve $x^2 + y^2 = 0$. After the blow-up $x\to xy$ it takes the form $y^2 (x^2+1) = 0$, i.e.\ $\alpha = 2$ and $\#\phi^{-1} = 2$. Thus, $\epsilon = 0$ in this case. 
Next, consider $x^2 + y^3 = 0$. The blow-up $x \to xy$ leads to $y^2 (x^2+y) = 0$, which means $\alpha = 2$ and $\#\phi^{-1} = 1$. Thus, $\epsilon = 1$.
Now, we are able to do the induction step. The curve $x^2+y^{n+2} = 0$ is blown up ($x\to xy$) to $y^2 (x^2+y^n) = 0$. Thus, $\alpha = 2$ and $\epsilon_{x^2+y^{n+2}} = 2 + \epsilon_{x^2+y^n}$, which proves the assertion.

\paragraph{${\bf x^3+y^n=0}$ for $n\geq 3$:}  

We prove again by induction but this time we need three initial steps. First, $x^3+y^3=0$ is blown up to $y^3(x^3+1)=0$ ($x\to xy$), $\alpha = 3$, $\# \phi^{-1} = 3$ and $\epsilon = 3 \cdot 2 - 3 = 3$. Second, $x^3+y^4=0$ is blown up to $y^3(x^3+y)=0$ ($x\to xy$), $\alpha = 3$, $\# \phi^{-1} = 1$ and $\epsilon = 3 \cdot 2 - 1 = 5$. 
Third, $x^3+y^5=0$ is blown up to $y^3(x^3+y^2)=0$ ($x\to xy$) and $\alpha = 3$. We have already shown that $\epsilon_{x^2 + y^3} = 1$.\footnote{Obviously, we are free to swap $x$ and $y$.} All in all, $\epsilon = 3 \cdot 2 + \epsilon_{x^2 + y^3} = 7$.

Finally, we perform the induction step $n \to n+3$. Consider the curve $x^3 + y ^{n+3} = 0$. After the blow-up $x \to xy$, the exceptional divisor $y^3$ factors out: $y^3(x^3+y^n) = 0$. Thus, $\epsilon_{x^3+y^{n+3}} = 3\cdot 2 + \epsilon_{x^3+y^n} = 6+2n-3 = 2(n+3) -3$ where we inserted the induction hypothesis $\epsilon_{x^3+y^n} = 2n-3$.

\subsection{Computation of  $\ki(X_{P_i})$}\label{sec:howToComputeChiP}
The last non-trivial element in formula~\eqref{eq:chi} is the topological Euler characteristic of the (partially resolved) fiber in $\hat Y_3$ over the enhancement points $P_i$. The (in general partial) resolution $\hat Y_3$ leads to several $\mathbb P^1$s intersecting each other in points. 
The intersection points are normal crossing singularities of the multi-component fiber. Since the Euler characteristic  of a single point is $1$, 
the standard prescription to compute the Euler characteristic of such a variety is as follows:  A smooth $\mathbb P^1$ has $\ki = 2$. Subtract one for every singular point on the $\mathbb P^1$ and sdd the contributions from all $\mathbb P^1$s up. Finally, add $+1$ for every singular point.

Let us make the prescription more concrete by considering some examples. The numbers in parenthesis denote the contributions to $\ki$. 
\begin{itemize} 
	\item The type III fiber. It has two components each of which has one singular point ($1+1$). In total there is one singular point (1). Thus, $\ki (\text{type III}) = 1+1+1 = 3$.
	\item The type IV fiber. It has three components all of which have a singular point ($1+1+1$). These three singular points are coincident (1). So, $\ki (\text{type IV}) = 3+1 = 4$. 
	\item The type $\mathrm I_0^*$ fiber. It has four components with one singular point ($1+1+1+1$), one component with four singular points ($2-4$) and all in all four singular points (4). Therefore, $\ki (\text{type I}_0^*) = 4-2+4 = 6$.
\end{itemize}

\subsection{Generalisation to Models With Three Discriminant Loci}  \label{cha:twoBranes}
Expression \eqref{eq:chi} applies to models with $\Delta  = \Sigma_0 \cup \Sigma_1$. In this appendix we generalize it to models with 
$\Delta  = \Sigma_0 \cup \Sigma_1 \cup \Sigma_2$ and identical gauge groups along $\Sigma_1$ and $\Sigma_2$, which will be studied further in appendix \ref{app:resolutionTwoBranes}.

Our starting points were  the four contributions to the topological Euler characteristic \eqref{eq:firstContribution} to \eqref{eq:fourthContribution}.
By assumption, the vanishing orders of $\Delta$ along $\Sigma_1$ and $\Sigma_2 $ are equal and are denoted again by $m$ and similarly for the vanishing orders  $\mu_f,\mu_g$.
Furthermore $B_1 = B_2$ for the number of enhancement points from intersection of $\Sigma_i$ with $\Sigma_0$.
 In addition there is a new type of enhancement locus  ($\{z_1 = 0\} \cap \{ z_2=0 \}$) called $R$. This gives an extra contribution to the topological Euler characteristic, $\chi_{top}(\pi^{-1}(R)) = 1 \cdot \chi_{top} (X_R)$.
 The second contribution \eqref{eq:secondContribution}, which took the topology of the brane without all enhancement points into account, becomes
  \begin{align*}
  \chi_{top} \left(\pi^{-1}\big((\Sigma_1 \cup \Sigma_2) \setminus P_1 \setminus P_2 \setminus R\big) \right) &= 2 \cdot \chi_{top} (\pi^{-1}( \Sigma_1 \setminus  P_1 \setminus  R))\\
  & =  2 \cdot m \cdot (2 - 2g-(B_1+1))
  \end{align*}
 The formula of the third contribution \eqref{eq:thirdContribution} relies on the fact that $\Sigma_0 \in -12 K_B -m \Sigma_1$ in the old situation. Here, $\Sigma_0 \in -12 K_B-m_1\Sigma_1 - m_2 \Sigma_2 = -12 K_B - 2m\Sigma_1$. Therefore, we have to replace $m \to 2m$.
 Similarly, one has to replace $\mu_f\to 2\mu_f$ and $\mu_g\to 2\mu_g$ in the formula for the number of cusps.
With the same argument as before the intersection multiplicity $\mu_i (f,g)$ can be set to zero.

Then, the contributions to $\chi_{top}(Y_3)$ are:
\begin{itemize}
  \item $\chi_{top} (\pi^{-1}(P_1 \cup P_2)) =2B_1\cdot \chi_{top} (X_{P_1}) $.
  
  \item $\chi_{top} (\pi^{-1}(R)) = \chi_{top} (X_R)$.
  
  \item $\chi_{top} (\pi^{-1}((\Sigma_1 \cup \Sigma_2) \setminus P_1 \setminus P_2 \setminus R)) =  2 m \cdot (2 -   2g-(B_1+1))$.
  
  \item $\chi_{top} (\pi^{-1} (\Sigma_0 \setminus  Q \setminus P_1 \setminus P_2)) = -11 \cdot 12 K_B^2 + 2mK_B \cdot \Sigma_1 + 4m^2 \Sigma_1^2+4m\Sigma_1\cdot \Sigma_0+ 2\epsilon_1 B_1+C$.
  
  \item $\chi_{top} (\pi^{-1} (Q)) =2C$ with the number of cusps $C = 24 K_B^2 + \Big(8\mu_g  + 12 \mu_f  \Big) K_B \cdot \Sigma_1 + 4\mu_f  \mu_g  \Sigma_1^2.$
\end{itemize}
All in all, $\ki$ is given by the more compact expression:
\bea\label{eq:chiTwoBranes}
\ki (Y_3) &=& -540 + \ki (X_R) + 2 \,B_1 \, \Big(\ki (X_{P_1}) + \epsilon_1\Big)+  m \cdot \Big(140 - 4m - 2B_1 \Big) - \nonumber \\
&&  -72 \mu_g  - 108 \mu_f   + 12 \mu_f  \mu_g.
\eea

\section{Toric Resolution of a Type IV Model}\label{app:toricResolutionsOneBrane}

In the main text we have encountered Weierstrass models with Kodaira fiber of type II and type III in codimension-one. Such models are interesting by themselves as they are inherently non-perturbative. The remaining model in this list of Kodaira outliers is Kodaira type IV, corresponding to a non-perturbative realisation of gauge  algebra $SU(3)$ (split) or $Sp(1)$ (non-split). 
In this appendix we describe the resolution of such a split type IV model, which, to the best of our knowledge, has not been presented in the literature before. 

A general type IV Weierstrass model is defined by arranging for vanishing orders of the following type along a divisor $\Sigma_1: z_1 = 0$,
\bea
f = z_1^{2} \, f_0, \qquad g = z_1^2 \, g_0 \qquad \rightarrow\quad \Delta = z_1^4 (27 g_0^2 + 4 f_0^3 z_1^2). 
\eea
The codimension-two enhancement points $P_1$ at $\Sigma_1 \cap \Sigma_0$ lie at 
\bea
P_1: z_1 = g_0 = 0  \quad {\rm with} \quad {\rm ord}(f,g,\Delta)|_{P_1} = (2,3,6),
\eea
corresponding to an enhancement from type IV to type $I_0^*$ in the fiber.
We realize a split version as a Tate model with vanishing orders 
\bea
a_i = \tilde a_i\,  z^{k_i}_i, \qquad k_i = (1,1,1,2,3) \quad {\rm for} \,\,  i=1,2,3,4,6 \,.
\eea
Similarly to the procedure described in Section \ref{IIIresolution} for the type III models, we resolve the singularities in the fiber by performing two blow-ups of the ambient space,
\begin{align}
x \to e_1\, e_2 \, x, \quad y \to e_1\, e_2^2\, y, \quad z_1 \to e_0\, e_1\, e_2 \, .
\end{align}
The proper transform $PT$ is now the non-generic hypersurface
\bea
PT = - e_1 x^3 + e_2 y^2 + \tilde a_1 e_0 e_1 e_2 x y z - \tilde a_2 e_0 e_1 x^2 z^2 + \tilde a_3 e_0 y z^3 -
  \tilde a _4 e_0^2 e_1 x z^4 - \tilde a_6 e_0^3 e_1 z^6
\eea
in the toric ambient space with scaling relations listed in table \ref{tab:ScalingRelationsIV}.
\begin{table}
\centering
\begin{tabular}{lccccccc}
\toprule
& $x$ & $y$ & $z$ & $e_0$ & $e_1$ & $e_2$ \\\midrule
$[Z]$ & 2&3&1&$\cdot$&$\cdot$ & $\cdot$\\
$[E_1]$&$-1$&$-1$&$\cdot$&$-1$&$1$ & $\cdot$\\
$[E_2]$&$-1$&$-2$&$\cdot$&$-1$&$\cdot$ & $1$\\\bottomrule
\end{tabular}
\caption{Toric weights for the fiber ambient space of the  type IV Tate model.}
\label{tab:ScalingRelationsIV}
\end{table}
This toric ambient space admits four different triangulations. For definiteness, consider the phase with Stanley-Reisner ideal
\begin{align}
{\rm SRI} = \langle ye_1,ze_1,ze_2,xyz,xye_0,xe_0e_2\rangle.
\end{align}
This time $PT$ is free of residual singularities and hence defines a crepant resolution to a smooth Calabi-Yau threefold $\hat Y_3$.\footnote{In fact, even for higher dimensional bases no singularities remain due to the SRI constraints.}

The fibral rational curves in codimension one, over $\Sigma_1$, are given by
\begin{align*}
\p_A:\: PT|_{e_0 \to 0} &= y^2 e_2-x^3 e_1,\\
\p_B:\: PT|_{e_1 \to 0} &= y\cdot (z^3 \tilde a_3 e_0+y e_2)\\
\p_C:\: PT|_{e_2 \to 0} &= \tilde a_3 e_0\, y z^3  - x^3 e_1-\tilde a_6 e_0^3 e_1 \, z^6 - \tilde a_4 e_0^2 e_1\, x z^4- \tilde a_2 e_0 e_1 \, x^2 z^2 \, .
\end{align*}
While the second equation factorises, the locus $\ib{e_1}\cap \ib y$ is forbidden by the SRI. Hence one finds altogether three curves intersecting in one point $\ib{e_0} \cap \ib{e_1} \cap \ib{e_2}$, as required for a type IV fiber. The curve $\p_A$ is intersected by the zero section $z=0$, while $\p_B$ and $\p_C$ are identified with the simple roots $\alpha_1$ and $\alpha_2$ of $SU(3)$, respectively. 
As always, the Lie algebra roots correspond to minus the intersection numbers with the resolution divisors representing the Cartan generators.
In this sense, the simple roots are given by
\bea
\alpha_1 = (2,-1) = (-E_1 \cdot \p_B, -E_2 \cdot \p_B), \qquad \alpha_1 = (-1,2) = (-E_1 \cdot \p_C, -E_2 \cdot \p_C).
\eea

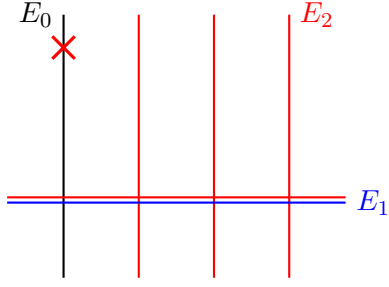
\begin{figure}
\centering

\begin{tikzpicture}[scale=1,thick]
\coordinate (a) at (1,0);
\coordinate (b) at (2,0);
\coordinate (c) at (3,0);
\coordinate (d) at (4,0);
\coordinate (len) at (0,3.5);
\coordinate (end) at (4.75,1);
\coordinate (shift) at (0,.07);
\draw (a) -- ($(a) + (len)$) node [very near end,draw,cross out,very thick, red] {};
\draw [red] (b) -- ($(b) + (len)$);
\draw [red] (c) -- ($(c) + (len)$);
\draw [red] (d) -- ($(d) + (len)$);
\draw [blue] (0.25,1) -- (end);
\draw [red] ($(0.25,1) + (shift)$) -- ($(end) + (shift)$);

\node [left] at ($(a) + (len)$) {$E_0$};
\node [right,blue] at (end) {$E_1$};
\node [right,red] at ($(d) + (len)$) {$E_2$};
\end{tikzpicture}

\caption{Affine Dynkin diagram of the resolved $\{z_1\}\cap \{\tilde a_3\}$ locus. The red cross denotes the intersection with the zero-section $z=0$ of the Weierstrass model. The blue and red colour indicates the splitting of $\p_{A,B,C}$.}
\label{fig:22z1a3}
\end{figure}

The next step is to take a look at the fiber enhancement in codimension two. From the discriminant 
\begin{align}
\Delta =  \tfrac1{16} \,z_1^4 \cdot \left( 27\, a_3^4 + \mathcal O (z_1)  \right)
\end{align}
one reads off that the fiber enhances at $z_1 = 0 =  \tilde a_3$, corresponding to the points $P_1$.
The fiber over this locus takes the form
\begin{align*}
PT|_{e_0 \to 0 ,\, \tilde a_3 \to 0} &= y^2e_2-x^3e_1, \\
PT|_{e_1 \to 0 ,\, \tilde a_3 \to 0} &= y^2 e_2,\\
PT|_{e_2 \to 0 ,\, \tilde a_3 \to 0} &= -e_1 \cdot (x^3 + \tilde a_2 x^2 (e_0 z^2) + \tilde a_4 x (e_0 z^2)^2 + \tilde a_6 (e_0 z^2)^3 )\\
&=  - e_1 \cdot (x-e_0 z^2 \cdot f_1(\tilde a_i)  )\cdot(x-e_0 z^2 \cdot f_2(\tilde a_i)  )\cdot (x-e_0 z^2 \cdot f_3(\tilde a_i)  ).
\end{align*}
Only the curve $\p_C$ splits into four components. If we denote the curves $\p_A$ and $\p_B$ over $z_1=\tilde a_3=0$  by $\p_a$ and $\p_b$,
then altogether 
\bea
\p_A \rightarrow \p_a, \qquad \p_B \rightarrow \p_b, \qquad \p_C \rightarrow \p_b \cup \p_c \cup \p_d \cup \p_
e.
\eea
We observe five distinct components, one with multiplicity two, intersecting as expected for a $\text I_0^*$ fiber, see figure~\ref{fig:22z1a3}. 
Computing $-1$ times the intersection numbers with the two resolution divisors $E_1$ and $E_2$, we find the $SU(3)$ weights of these curves to be
\begin{align*}
\p_a : (-1,-1), \quad \p_{b} : (2,-1), \quad \p_{c,d,e} : (-1,1).
\end{align*}
The highest weight of the fundamental representation is $w_1 = (1,0)$. There are three possibilities to represent the highest weight in terms of holomorphic fibral curves,
\begin{align*}
(\p_b + \p_c): (1,0), \quad (\p_b + \p_d): (1,0), \quad (\p_b + \p_e): (1,0).
\end{align*}
To construct the other states, we have to act with the simple roots on the highest weight vector. For example, for starting with $\p_b + \p_c$ this gives
\begin{align*}
w_1 :&  \quad  \p_b + \p_c,\\
w_1 -\alpha_1 :& \quad  (\p_b + \p_c) - \p_b = \p_c,\\
w_1 - \alpha_1 - \alpha_2 :& \quad \p_c -( \p_b + \p_c + \p_d + \p_e) =  - (\p_b + \p_d + \p_e).
\end{align*}
Each these three curves can be wrapped by an M2-brane as well as an anti-M2-brane. This gives rise to one hypermultiplet in the ${\bf 3}$ of SU(3).
Repeating this for the remaining two highest weights, we conclude that we find in total {\it three copies of the fundamental representation} in the resolved fiber at each enhancement point $P_1$.
This spectrum is in full agreement with the cancellation of both gravitational and SU(3) gauge anomalies. 
For the special case of a base $B_2 = \mathbb P^2$ we list the charged and uncharged spectrum, computed via $\chi_{\rm top}(\hat Y_3)$, in table \ref{tab:oneBraneResults-smooth}.

\begin{table}
\centering
\begin{tabular}{lrrrr}
\toprule
$(\mu_f ,\mu_g)$    & $(0,0)$ & \hspace{3em}$(1,2)$ &\hspace{3em} $(1,3)$ & \hspace{3em}$(2,2)$\\
$m$                 & ---     & $3$     & $3$     & 4\\\midrule
Enhancements        & ---     & III $\to$ IV & III $\to \text I_0^*$ & IV $\to \text I_0^*$\\
Gauge Group         & ---     & $SU(2)$ & $SU(2)$ & $SU(3)$\\
$n_V=\dim (G)$      & ---     & $3$     & $3$     & $8$\\
$\mathrm{rk}(G)$    & ---     & $1$     & $1$     & $2$\\
$h^{1,1} (\tilde Y_3)$ & $2$  & $3$     & $3$     & $4$\\
$B_1$               & $0$     & $11$    & $11$    & $8$\\
$\ki(X_{P_1})$      & ---     & $4$     & $3$     & $6$\\
$\epsilon_1$        & ---     & $-1$    & $3$     & $2$\\\midrule
$\ki (\tilde Y_3)$         & $-540$  & $-456$  & $-456$  & $-408$\\\midrule
non-loc neutral hypers $n^0_{H,n-l}$ & $273$& $232$ & $232$ & $209$\\
charged hypers $n^c_{H}$& $0$& $44$    & $44$    & $72$\\
Representation      & ---     & $2\times$fund.& $2\times$fund.& $3\times$fund.\\\midrule
$273-(n^0_{H}+n^c_{H}-n_V)$ & 0   & 0    & 0    & 0\\\bottomrule
\end{tabular}
\caption{Smooth Weierstrass models over $B_2 = \mathbb P^2$. The first column corresponds to a generic Weierstrass model, the second and third column correspond to the smooth type III models discussed in  Section \ref{sec_nontrivialgaugegroup}. }
\label{tab:oneBraneResults-smooth}
\end{table}

\section{Models With $\Delta = \Sigma_0 \cup \Sigma_1 \cup \Sigma_2$}\label{app:resolutionTwoBranes}

\begin{table}[h]
\centering
\begin{tabular}{lrrrr}
\toprule
$(\mu_f ,\mu_g)$    & $(1,2)$ & \hspace{2em}$(1,3)$ &\hspace{2em} $(1,4)$&\hspace{2em} $(1,5)$\\\midrule
$\ki(X_{P_1})$      & 4&3&3&3\\
$\ki(X_{R})$        &4&4&4&4 \\
$\epsilon_1$        &$-1$ & 3&7&11\\\midrule
$a$                 & -- & -- & 2&3\\
$m_P$               &  0&0&1&2\\
$\ki (\tilde Y_3)$         & $-380$&$-380$&$-360$&$-340$\\\midrule
non-loc neutral hypers $n^0_{H,n-l}$& 195&195&175&155\\
localisaed neutral hypers $n^0_{H}$& --&--&20 &40 \\\midrule
charged hypers $n^c_{H}$&84&84&84&84\\\midrule
$273-(n^0_{H,n-l}+n^0_{H,l}+n^c_{H}-n_V)$ & 0   & 0    & 0  &0 \\\bottomrule
\end{tabular}
\caption{Type III $\times$ Type III Models over $B_2 = \mathbb P^2$.}
\label{tab:twoBraneResultsnonResGauge}
\end{table}

In this appendix, the analysis of F-theory models with one brane is generalized to models with two identical branes in generic position to each other, i.e.\ models with two identical gauge group factors. This provides a richer structure of singularity types and geometry of the fibration. 
We will consider models with two identical branes wrapped on the divisors $\Sigma_1: z_1=0$ and $\Sigma_2: z_2=0$. These models are defined via the requirement that $f$ and $g$ in the Weierstrass model vanish to certain orders along the divisors. Since the two branes shall be identical we need two numbers to specify a model. We employ the following notation: The model with $f = (z_1z_2)^{\mu_f} f_0$ and $g = (z_1z_2)^{\mu_g} g_0$ is called \emph{$[\mu_f\mu_g]$-model.}

\subsection{The $[n1]$-Models: Type II $\times$ Type II} \label{[n1]models}
Let us start out with the $[n1]$-models generalizing the geometries discussed in Section \ref{typeIImodel} by taking
\begin{align}
f = (z_1z_2)^n f_0, \qquad g = (z_1z_2) g_0 \qquad \Longrightarrow \quad 
\Delta = \underbrace{z_1^2}_{\Sigma_1} \underbrace{z_2^2}_{\Sigma_2} \underbrace{(27 g_0^2+4 (z_1 z_2)^{3n-2} \,f_0^3)}_{\Sigma_0}.
\end{align}
If we specialise these models to base $B_2 = \mathbb P^2$, where $f\in \mathcal O (12)$, and identify $z_1$ and $z_2$ with two of the homogenous coordinates $[z_0 : z_1 : z_2]$, the allowed range for  $n$ is $1 \leq n \leq 6$. 

The fiber over $\Sigma_1$ and $\Sigma_2$ is type II with trivial gauge group. 
As a novelty compared to the single brane models, we encounter now an intersection of two type II components at $\Sigma_1 \cap \Sigma_2$. This locus is  called $R$ in the notation of appendix \ref{cha:twoBranes}, and characterized as follows:
\bea
R: z_1 = z_2= 0, \qquad  {\rm ord}(f,g,\Delta) =(2n,2,4)  \qquad \Longrightarrow  \quad {\rm II} \, {\rm x} \,  {\rm II} \rightarrow {\rm IV} \, .
\eea
In addition, over the $B_i=16$ points of type $P_i$  where $z_i = g_0 = 0$, corresponding to the intersection of $\Sigma_i$ with the residual discriminant $\Sigma_0$, the fiber enhances as follows:
\bea
&n=1:  \qquad B_i:  z_i = g_0 = 0, \qquad {\rm ord}(f,g,\Delta) =(1,2,3) \qquad &{\rm II} \,  \rightarrow \, {\rm  III}, \\
&n\geq2:  \qquad B_i:  z_i = g_0 = 0, \qquad {\rm ord}(f,g,\Delta) =(n,2,4) \qquad &{\rm II} \,  \rightarrow \, {\rm IV}\, .
\eea
Both the type IV enhancements over $R$ and $B_i$ (for $n \geq 2$) and   and the type III enhancements over $B_i$ (for $n=1$) are isolated terminal Kleinan singularities of local form $z^3+x_1^2+x_2^2+x_3^2=0$, with Milnor number $m_{P_i} = m_R = 2$. Hence each of these enhancement points carries $2$ massless uncharged hypermultiplets (independently of the choice of base).

To compute the Euler characteristic \eqref{eq:chiTwoBranes} we set $\ki(X_{P_i}) = 2 = \ki (X_R)$, corresponding to the singular non-resolved fiber. 
Over base $B_2 = \mathbb P^2$ and with these values, the general expression \eqref{eq:chiTwoBranes} simplifies to
\begin{align*}
\ki (Y_3) = -346 - 96 \,n + 32 \,\epsilon.
\end{align*}
The remaining task is to determine $\epsilon$. The curve $\Sigma_0$ takes the form $g_0^2+z_{1,2}^{3n-2} = 0 $ near an intersection locus with $\Sigma_{1}$ or $\Sigma_2$. Hence, $\epsilon = 3n-4$ (see \eqref{eq:epsilonFormulae}).  This cancels the $n$-dependence of $\ki (Y_3)$ and we end up with the value $\ki ( Y_3)= -474$ for all $n$. 
From equation~\eqref{eq:cd} we obtain $239+\frac12 \sum m_P = 272$ complex structure deformations and hence 273 uncharged massless hypermultiplets, as required by anomaly cancellation.

\begin{table}
\centering
\begin{tabular}{lccccccc}
\toprule
& $x$ & $y$ & $z$ & $e_0$ & $e_1$ & $f_0$ & $f_1$\\\midrule
$[Z]$ & 2&3&1&$\cdot$&$\cdot$&$\cdot$&$\cdot$\\
$[E_1]$&$-1$&$-1$&$\cdot$&$-1$&$1$&$\cdot$&$\cdot$\\
$[F_1]$&$-1$&$-1$&$\cdot$&$\cdot$&$\cdot$&$-1$&$1$\\\bottomrule
\end{tabular}
\caption{Scaling Relations for the fiber ambient space of the resolved $[1n]$-models.}
\label{tab:scalings1n}
\end{table}
One possible choice for the Stanley-Reisner ideal is:
\begin{align*}
\langle ze_1, zf_1, e_0f_1, xyz, xye_0, xyf_0 \rangle.
\end{align*}

\subsection{The $[1n]$-Models ($n>1$): Type III $\times$ Type III}\label{sec:1nModels}

These models generalize the geometries analyzed in Section \ref{sec_typeIIImodel} in the sense that
\begin{align}\label{eq:Delta1n}
{\rm ord}(f,g,\Delta)|_{z_i} = (1,n,3)   \quad \Longrightarrow \quad \Delta = z_1^3z_2^3 \,\Big(4f_0^3 + 27\, (z_1z_2)^{2n-3}\, g_0^2\Big)
\end{align}
with $n >1$, corresponding to gauge group $G = SU(2) \times SU(2)$. 
The intersection of both 7-branes at the point R,
\bea
R: z_1 = z_2= 0, \qquad  {\rm ord}(f,g,\Delta) =(2,2n,6)  \qquad \Longrightarrow  \quad {\rm III} \, {\rm x} \,  {\rm III} \rightarrow {\rm I}_0^\ast \, ,
\eea
is new compared to the models in Section \ref{sec_typeIIImodel}, whereas the $B_i$ loci $\Sigma_i = f_0 = 0$ behave as in eq. (\ref{TypeIIImodelordersP1}).
The explicit resolution of a Tate model realisation of these fibrations given below confirms the structure of the resolved fiber over the locus $R$ as a monodromy reduced I$^*_0$ fiber with $\chi_{\rm top}(X_R)=4$. This fiber carries one massless hypermultiplet in representation $({\bf 2},{\bf 2})$ of $G = SU(2) \times SU(2)$. Furthermore, as in Section \ref{sec_typeIIImodel}, for $n \geq 4$ the fiber over the points $B_i$ exhibit a residual Kleinan singularity of the form (\ref{RHMtypeIII1}) - (\ref{localhyperIII4}), which is responsible for the localisation of a corresponding number of uncharged hypermuliplets at these points in addition to the two hypermultiplets in the $({\bf 2},1)$ or $(1,{\bf 2})$, respectively. It can readily be checked that this charged spectrum leads to an anomaly free spectrum. Furthermore, over base $B_2 = \mathbb P^2$, we can compute the topological Euler characteristic via \eqref{eq:chiTwoBranes} (see \, table \,\ref{tab:twoBraneResultsnonResGauge}). In this case, the number of points $B_i$ is $10$, and the details of the computation parallel the analysis in Section \ref{sec_typeIIImodel}.

\begin{table}
\centering
\begin{tabular}{llr}\toprule
Model & Singular locus after resolution & $a$\\\midrule
$[12]$ & $\varnothing$ & --\\
$[13]$ & $\varnothing$ & --\\
$[14]$ & $\big( \ib{e_1} \cap \ib{a_4} \cap \ib x \cap \ib y \big) \cup \big( \ib{f_1} \cap \ib{a_4} \cap \ib x \cap \ib y \big)$ & 3\\
$[15]$ & $\big( \ib{e_1} \cap \ib{a_4} \cap \ib x \cap \ib y \big) \cup \big( \ib{f_1} \cap \ib{a_4} \cap \ib x \cap \ib y \big)$& 3\\\bottomrule
\end{tabular}
\caption{Singular locus of proper transforms. The singularity parameter $a$ describes the local form of the singularity: $z^a + x_1^2+x_2^2+x_3^2$.}
\label{tab:singLociTwoBrane}
\end{table}

We resolve the $[1n]$-models by realising them in Tate form, choosing the Tate vanishing orders listed in table~\ref{tab:vanishingOrders} along $z_1=0$ and $z_2=0$. Let us explicitly consider the cases  $n = 2,3,4,5$ since all conceptional features appear already here.
The singularities at $z_i = x = y =0$  are resolved by the two blow-ups 
\begin{align}\label{eq:blowupOneTime}
x \to e_1f_1x, \quad y \to e_1f_1y, \quad z_1 \to e_0e_1,\quad z_2 \to f_0f_1,
\end{align}
which lead to the proper transform $PT$ of the Tate form 
\begin{align*}
PT =& -e_1 f_1 \, x^3 + y^2 + \tilde a_1\, e_0^n e_1^n f_0^n f_1^n\,  x y z - 
 \tilde a_2\,  e_0^n e_1^n f_0^n f_1^n\,  x^2 z^2 + \\
& + \tilde a_3\,  e_0^n e_1^{n-1} f_0^n f_1^{n-1} \, y z^3 - \tilde a_4\,  e_0 f_0 \, x z^4 - \tilde a_6 \, e_0^n e_1^{n-2} f_0^n f_1^{n-2} \, z^6 \quad \text{for } n \geq 2. 
\end{align*}
Again, we view $PT$ as a hypersurface in a toric fiber ambient space with coordinates $x$, $y$, $z$, $e_1$, $f_1$, $e_0$, $f_0$. The associated toric weights are displayed in table~\ref{tab:scalings1n}.

\begin{figure}
\centering


\begin{tikzpicture}[thick]
\coordinate (a) at (1,0);
\coordinate (b) at (2,0);
\coordinate (c) at (3,0);
\coordinate (d) at (4,0);
\coordinate (len) at (0,3.5);
\coordinate (end) at (4.75,1);
\coordinate (shift) at (0,.07);

\draw (a) node [below] {$\mathbb P^1_a$} -- ($(a) + (len)$) node [very near end,draw,cross out,very thick, red] {};
\draw (b) node [below] {$\mathbb P^1_c$} -- ($(b) + (len)$);
\draw [dashed] (c) -- ($(c) + (len)$);
\draw [dashed] (d) -- ($(d) + (len)$);
\draw (0.25,1) -- (end) node [right] {$\mathbb P^1_{b_{1,2}}$};
\draw ($(0.25,1) + (shift)$) -- ($(end) + (shift)$);

\end{tikzpicture}

\caption{Affine Dynkin diagram of the resolved $\{z_1\}\cap \{z_2\}$ locus. The red cross denotes the intersection with the zero-section $z=0$ of the Weierstrass model.}
\label{fig:1nz1z2}
\end{figure}
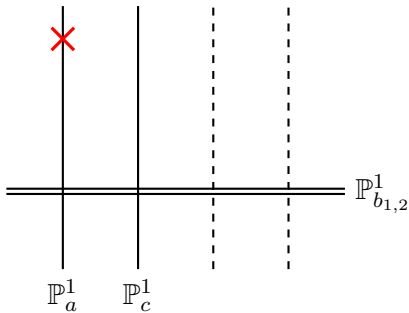

In complete analogy to the case studied in Section \ref{IIIresolution}, for $n \geq 4$ there remain singularities in codimension 2 as listed in table \ref{tab:singLociTwoBrane}.
These residual singularities result in uncharged localised hypers.
Indeed, away from the point $z_1 = z_2=0$ the structure of fibers is identical to the pattern for a single type III brane detailed in Section  \ref{IIIresolution}.
Hence it only remains to analyze the fiber type at the intersection of the two type III singularities at $R: z_1 = z_2 =0$. 
Na\"ive application of  Kodaira's classification predicts an $\text I_0^*$ fiber. Direct inspection reveals the following fibral curves over $R$ (note that $e_0 \to 0, f_1\to 0$ is forbidden by the SRI),
\begin{align*}
\p_a:\: PT|_{e_0 \to 0,\, f_0\to 0} &= -e_1 f_1 x^3+y^2, \\
\p_{b_{1,2}}:\: PT|_{e_1 \to 0,\, f_0\to 0} &= y^2, \\
\p_c:\: PT|_{e_1 \to 0,\, f_1\to 0} &= \begin{cases}
y^2-a_4 e_0 f_0 x z^4-a_6 e_0^2 f_0^2 z^6\quad & \text{for } a=2,\\
y^2-a_4 e_0 f_0 x z^4\quad & \text{for } a>2.
\end{cases}
\end{align*}
The second line indicates the presence of two copies of the rational curve $ e_1 = f_0= y = 0$. These two copies will be denoted by $\p_{b_{i}}$, $i=1,2$.
We can interpret this fiber as a monodromy reduced $\text I_0^*$ fiber with two nodes deleted, where $\p_{b_{i}}$, $i=1,2$ corresponds to the middle $\mathbb P^1$ of multiplicity 2, intersecting the two other curves once (see figure~\ref{fig:1nz1z2}).

The negative of the intersection numbers with the resolution divisors $E_1$ and $F_1$ give the $U(1) \times U(1)$ Cartan charges
\bea
\p_{b_i} =(1,-1), \qquad \p_c: (0,2) \,,
\eea
while $\p_a$ is intersected by the zero section and hence plays no role in determining the weight lattice. 
If we denote the $w_0^{E/F}$ the highest weight of the fundamental representation of the two $SU(2)$ factors, and by $\alpha_01^{E/F}$ their simple root,
we can make the following identification between holomorphic and anti-holomorphic curves in the fiber and the weights of a bifundamental representation $({\bf 2}, {\bf 2})$,
\begin{align*}
\begin{pmatrix}
(w_0^E+\alpha_1^E,w_0^F)\\
(w_0^E,w_0^F)\\
(w_0^E,w_0^F+\alpha_1^F)\\
(w_0^E+\alpha_1^E,w_0^F+\alpha_1^F)\\
\end{pmatrix}
=
\begin{pmatrix}
\p_{b_1}\\
-(\p_{b_2} + \p_c)\\
-\p_{b_2}\\
\p_{b_1} + \p_c
\end{pmatrix}
=
\begin{pmatrix}
(1,-1)\\
(-1,-1)\\
(-1,1)\\
(1,1)
\end{pmatrix}.
\end{align*}
Here we explicitly distinguish between the two curves $\p_{b_1}$ and $\p_{b_2}$.
M2 branes wrapping each of these four curves with positive and negative orientation give rise to a full hypermultiplet in the $({\bf 2}, {\bf 2})$ of $SU(2) \times SU(2)$. This perfectly matches the prediction from the anomaly constraints.

\section{Summary of All Models With $\Delta = \Sigma_0 \cup \Sigma_1$}

The results for all models in this paper with enhancements along a single divisor on $\mathbb P^2$ are summarized in table \ref{tab:summaryOneBraneModels}.

\begin{table}
\centering
\begin{tabular}{rrrllrrrrrrrr}
\toprule
& $(\mu_f, \mu_g)$& $m$& Enh. & Gauge Gr. & $n_V$ & $\epsilon_1$ & $\ki (\tilde Y_3)$ & $n_{H,n-l}^0$ & $B_1$ & $m_P$ & $n_{H,l}^0$ & $n_H^c$\\\midrule
\parbox[t]{2mm}{\multirow{4}{*}{\rotatebox[origin=c]{90}{Table 8}}} & $(0,0)$ &0& ---&---&---&---&$-540$&$273$&---&---&---&---\\
& $(1,2)$ & 3& III $\to$ IV & $SU(2)$ & 3 & $-1$ & $-456$& $232$ & $11$ & ---&---&$44$\\
& $(1,3)$ &  3& III $\to \mathrm I_0^*$  & $SU(2)$ & 3 & $3$ & $-456$& $232$ & $11$ & ---&---&$44$\\
& $(2,2)$ &  4& IV $\to \mathrm I_0^*$ & $SU(3)$ & 8 & $2$ & $-408$& $209$ & $8$ & ---&---&$72$\\\midrule
\parbox[t]{2mm}{\multirow{2}{*}{\rotatebox[origin=c]{90}{\footnotesize Table 2}}}& $(1,1)$ & 2 & II $\to$ III & --- & --- & $-1$ & $-506$& $239$ & $17$ & 2 & $34$ & ---\\
& $(2,1)$ & 2 & II $\to$ IV & --- & --- & $2$ & $-506$& $239$ & $17$ & 2 & $34$ & ---\\\midrule
\parbox[t]{2mm}{\multirow{3}{*}{\rotatebox[origin=c]{90}{Table 3}}} & $(1,4)$ & 3&III $\to \mathrm I_0^*$ & SU(2) & 3 & 7 & $-445$ & $221$ & $11$ & 1 & $11$ & $44$\\
& $(1,5)$ & 3&III $\to \mathrm I_0^*$ & SU(2) & 3 & $11$ & $-434$ & $210$ & $11$ & 2 & $22$ & $44$\\
& $(1,7)$ & 3&III $\to \mathrm I_0^*$ & SU(2) & 3 & $19$ & $-412$ & $188$ & $11$ & 4 & $44$ & $44$\\\bottomrule
\end{tabular}
\caption{Models over $B_2 = \mathbb P^2$ with enhancements over a single divisor.}
\label{tab:summaryOneBraneModels}
\end{table}




\bibliography{Fsing}{}
\bibliographystyle{JHEP}


\end{document}